\documentclass[superscriptaddress,aps,preprintnumbers,amsmath,amssymb,,sort&compress,nofootinbib,english,notitlepage]{revtex4-1}

\usepackage{amsfonts,amsmath,amssymb,bm,tensor}
\usepackage{comment}
\usepackage{ifpdf}
\usepackage{slashed}
\usepackage{color}
\usepackage[mathscr]{eucal}
\usepackage[utf8]{inputenc}

\usepackage{cancel}

\ifpdf
  \usepackage{graphicx}     
  \usepackage[bookmarksopen,colorlinks=true,linkcolor=bblue,citecolor=ppink,urlcolor=ppink]{hyperref}
\else     
  \usepackage[dvipdfmx]{graphicx}     
  \usepackage[dvipdfmx,bookmarksopen,colorlinks=true,linkcolor=bblue,citecolor=ppink,urlcolor=darkred]{hyperref}
\fi

\definecolor{red}{rgb}{1,0,0}
\definecolor{darkred}{rgb}{0.6,0,0}
\definecolor{darkgreen}{rgb}{0.992447,0.623778,0.034597}
\definecolor{ppink}{rgb}{1,0.4,0.4}
\definecolor{bblue}{rgb}{0.284602,0.317763,0.963947}

\newcommand{\1}{\mbox{1}\hspace{-0.25em}\mbox{l}}
\newcommand{\Red}[1]{{\color{red} {#1}}}

\newcommand{\vev}[1]{ \left< {#1} \right> }

\newcommand{\prn}[1]{\left( {#1} \right)}

\newcommand{\TODO}[1]{{\color{red}{$[[ \clubsuit\clubsuit$ \bf #1 $\clubsuit\clubsuit ]]$}}}

\newcommand{\bvec}[1]{\mbox{\boldmath $#1$}}

\makeatletter
\newcommand\footnoteref[1]{\protected@xdef\@thefnmark{\ref{#1}}\@footnotemark}
\makeatother

\begin{document}


\title{
Primordial Black Holes for the LIGO Events in the Axion-like Curvaton Model
}
\author{Kenta Ando}
\affiliation{ICRR, University of Tokyo, Kashiwa, 277-8582, Japan}
\affiliation{Kavli IPMU (WPI), UTIAS, University of Tokyo, Kashiwa, 277-8583, Japan}
\author{Keisuke Inomata}
\affiliation{ICRR, University of Tokyo, Kashiwa, 277-8582, Japan}
\affiliation{Kavli IPMU (WPI), UTIAS, University of Tokyo, Kashiwa, 277-8583, Japan}
\author{Masahiro Kawasaki}
\affiliation{ICRR, University of Tokyo, Kashiwa, 277-8582, Japan}
\affiliation{Kavli IPMU (WPI), UTIAS, University of Tokyo, Kashiwa, 277-8583, Japan}
\author{Kyohei Mukaida}
\affiliation{DESY, Notkestra{\ss}e 85, D-22607 Hamburg, Germany}
\author{Tsutomu T.~Yanagida}
\affiliation{Kavli IPMU (WPI), UTIAS, University of Tokyo, Kashiwa, 277-8583, Japan}
\affiliation{Hamamatsu Professor}

\begin{abstract}
\noindent
We revise primordial black hole (PBH) formation in the axion-like curvaton model and investigate whether PBHs formed in this model can be the origin of the gravtitational wave (GW) signals detected by the Advanced LIGO. In this model, small-scale curvature perturbations with large amplitude are generated, which is essential for PBH formation. On the other hand, large curvature perturbations also become a source of primordial GWs by their second-order effects. Severe constraints are imposed on such GWs by pulsar timing array (PTA) experiments. We also check the consistency of the model with these constraints. 
In this analysis,  it is important to take into account the effect of non-Gaussianity, which is generated easily in the curvaton model.
We see that, if there are non-Gaussianities, the fixed amount of PBHs can be produced with a smaller amplitude of the primordial power spectrum.
\end{abstract}

\date{\today}
\maketitle

\preprint{IPMU 17-0165}
\preprint{DESY 17-209}

\tableofcontents

\section{Introduction}

The primordial black hole (PBH) is one of the hot topics in cosmology and astrophysics.
It has been especially motivated in recent years by outstanding observational progress.
In 2015, the Advanced LIGO detected gravitational waves (GWs) directly for the first time~\cite{Abbott:2016blz}. 
This event, GW150914, comes from the merger of black holes (BHs) with $36.2^{+5.2}_{-3.8}M_\odot$ and $29.1^{+3.7}_{-4.4}M_\odot$ each. 
Another event, GW170104, comes from BHs with $31.2^{+8.4}_{-6.0}M_\odot$ and $19.4^{+5.3}_{-5.9}M_\odot$ ~\cite{Abbott:2017vtc}. 
Furthermore, the recently observed event, GW170814, comes from BHs with $30.5^{+5.7}_{-3.0}M_\odot$ and $25.3^{+2.8}_{-4.2}M_\odot$ each~\cite{Abbott:2017oio}.
These BHs are massive compared with $\sim10M_\odot$ BHs which have been observed via X-ray. 
The identity of such BHs is one of the outstanding problems.

One of the candidates is primordial black holes (PBHs)~\cite{
Bird:2016dcv,Clesse:2016vqa,
Sasaki:2016jop,
Eroshenko:2016hmn,
Carr:2016drx}. 
PBHs are BHs formed in the early stage of the universe by the gravitational collapse of the over-dense regions~\cite{Hawking:1971ei,Carr:1974nx,Carr:1975qj}. 
They can take a much broader range of mass than BHs formed at the end of stellar evolution. 
In this paper, we consider PBHs formed only in the radiation-dominated universe.\footnote{
It is suggested that the production rate of solar mass PBHs is enhanced by at least 2 orders of magnitude due to the QCD phase transition
\cite{Byrnes:2018clq}.
The shape of the spectrum in Fig.~\ref{fig:mass_spectrum} would not be affected so much since the spectrum is very sharp and it is sufficiently damped at solar mass. 
}
In order for PBHs to be formed, large primordial density perturbations on small scales are needed. 
However, large-scale perturbations are determined with good precision by the cosmic microwave background (CMB) observation and the amplitude is known to be small like $\mathcal P_\zeta\sim10^{-9}$~\cite{Ade:2015xua}.
Moreover, there are severe constraints on primordial perturbations for the scales which are larger than and close to the scales associated with $\sim30M_\odot$ PBHs:
the CMB $\mu$-distortion ~\cite{Kohri:2014lza,Fixsen:1996nj} and the change of big-bang nucleosynthesis (BBN)~\cite{Inomata:2016uip,Jeong:2014gna, Nakama:2014vla}.
Therefore, an extremely blue-tilted spectrum of primordial perturbations is required for the formation of $\sim30M_\odot$ PBHs.

The axion-like curvaton model~\cite{Kawasaki:2012wr, Kasuya:2009up} is known to be a model in which highly blue-tilted (iso)curvature perturbations can be generated. 
A curvaton generally indicates a field which generates curvature perturbations instead of the inflaton~\cite{Enqvist:2001zp, Lyth:2001nq, Moroi:2001ct}. 
In the context of PBH formation, the inflaton contributes to the large-scale perturbations and reproduces the CMB spectrum while the curvaton contributes to the small-scale perturbations with large amplitude. 
In Ref.~\cite{Kawasaki:2012wr}, PBH abundance is evaluated in terms of curvature perturbations. 
However, it has been recently suggested that the evaluation in terms of density perturbations should be more appropriate~\cite{Young:2014ana}.
Hence, we reconsider PBH formation in this model.

If there are large curvature perturbations, their second-order effects can become a dominant source of primordial GWs~\cite{Ananda:2006af,Baumann:2007zm,Saito:2008jc,Saito:2009jt,Bugaev:2009zh,Bugaev:2010bb}.
In the frequency region associated with $\sim M_\odot$, the amplitude of GWs is constrained severely by pulsar timing array (PTA) experiments~\cite{Inomata:2016rbd, Orlofsky:2016vbd}. 
One has to verify that the model does not contradict with the PTA experiments.
Naively speaking, the GW spectrum traces the curvature perturbation spectrum. 
In the double inflation model, which also generates blue-tilted curvature perturbations, the curvature perturbation spectrum has a red-tilted region and it is a key point to avoid the PTA constraints~\cite{Inomata:2016rbd}. 
On the other hand, in the axion-like curvaton model, the shape of the curvature perturbation at a given time is plateau-like on small scales and hence the PTA constraint seems more severe. 
However, in the curvaton model, a large non-Gaussianity can be generated. 
In such cases, the amplitude of GWs is suppressed when a given abundance of PBHs is explained~\cite{Young:2013oia,Nakama:2016gzw}.

In this paper, we revisit the axion-like curvaton model concerning the PBH formation and non-Gaussianity.
Then, we investigate the possibility that PBHs formed in this model explain the LIGO events. Also, we check the consistency with the PTA experiments taking account of the effect that the induced GWs are suppressed because of non-Gaussianity. 
In Sec.~\ref{axion_curvaton_model}, we explain the axion-like curvaton model and curvature perturbations generated there.
In Sec.~\ref{PBH_formation}, we review PBH formation.
In Sec.~\ref{sec_GW}, we review GWs coming from the second-order effects.
In Sec.~\ref{sec_Result}, we show that the axion-like curvaton model can explain the LIGO events.
In Sec.~\ref{sec_Conclusion}, we give a conclusion.


\section{Axion-like curvaton Model}\label{axion_curvaton_model}
In this section, we describe the essence of the axion-like curvaton model mainly following Ref.~\cite{Kawasaki:2012wr}. We will see that this model can generate a blue-tilted curvature perturbation spectrum.
We do not specify the inflation model.
The energy scale $H_\mathrm{inf}$ is only relevant and we neglect its variation during inflation.
We assume, as is common, that the oscillating inflaton behaves as non-relativistic matter after inflation until reheating.


\subsection{Potential}
The axion-like curvaton model is formulated in the framework of supersymmetry. The superpotential is given by
\begin{equation}
	W=hS(\Phi\bar\Phi-f^2),\label{superpotential}
\end{equation}
where $\Phi$, $\bar \Phi$ and $S$ are chiral superfields, $f$ is a certain energy scale, and  $h$ is a dimensionless coupling constant. There is a global $U(1)$ symmetry, and $\Phi$, $\bar \Phi$ and $S$ are assigned the charge $+1$, $-1$ and $0$ respectively. The scalar potential in the case of the global SUSY is derived from Eq.~\eqref{superpotential} as
\begin{equation}
	V=h^2|\Phi\bar\Phi-f^2|^2+h^2|S|^2(|\Phi|^2+|\bar\Phi|^2),
\end{equation}
where the scalar component of the superfield is denoted by the same symbol as the superfield.\\
\indent The potential has a flat direction
\begin{equation}
	\Phi\bar\Phi=f^2,\quad S=0.
\end{equation}
From now on, we assume that the field values always satisfy this condition. 
Taking account of the supergravity effect, the Hubble-induced mass terms
\begin{equation}
	V_H=c_1H^2|\Phi|^2+c_2H^2|\bar\Phi|^2+c_SH^2|S|^2
\end{equation}
lift the flat direction.\footnote{
There are also the soft SUSY breaking terms, but they are negligible during the inflation.
}
Here, $c_1$, $c_2$ and $c_S$ is a dimensionless constant whose value is in the order of unity. The minimum is determined by $|\Phi|\simeq|\bar\Phi|\simeq f.$
Without loss of generality, we can take the initial field values like $|\Phi|\gg|\bar\Phi|$, and then only one complex scalar field is relevant in the early stage of the universe~\cite{Kasuya:1996ns}. It is denoted as 
\begin{equation}
	\Phi=\frac{1}{\sqrt2}\varphi\exp\left(i\frac{\sigma}{f}\right).
\end{equation}
In this model, the field corresponding to the phase direction, $\sigma$, works as a curvaton, and the evolution of the radial direction, $\varphi$, makes the spectrum extremely blue tilted.

The potential of the relevant field is given by
\begin{equation}
	V_\varphi=\frac{1}{2}cH^2\varphi^2\label{SUGRA}.
\end{equation}
The field $\varphi$ is initially displaced far away from the minimum and rolls down to $\varphi_{\mathrm{min}}\simeq f$ during the inflation, which is essential for a blue-tilted curvature perturbation spectrum as shown later.
We assume that the global $U(1)$ symmetry is broken by some non-perturbative effect and that $\sigma$ obtains the following potential like the axion:
\begin{equation}
	V_\sigma=\Lambda^4\left[1-\cos\left(\frac{\sigma}{f}\right)\right]\simeq\frac{1}{2}m_\sigma^2\sigma^2,\label{curvaton_potential}
\end{equation}
where the curvaton mass is represented as $m_\sigma=\Lambda^2/f$ and the last equality holds when $\sigma/f$ is small. Hereafter, we assume that this approximation is always valid. 
The origin of $\sigma$ is determined to be the minimum of the potential.
The phase $\sigma/f$ takes an initial value at random, which is called the misalignment mechanism.
We denote the misalignment angle as $\theta$ ($=\sigma_i/f$).


\subsection{Dynamics}
We explain the dynamics of the curvaton in this subsection.
First, when the Hubble parameter becomes comparable to the curvaton mass $m_\sigma$, the curvaton field starts to oscillate with the initial amplitude $\sigma_i= f\theta$. We denote the ratio of the energy density of the curvaton to that of the radiation as $r$:
\begin{equation}
	r(\eta)=\frac{\rho_\sigma(\eta)}{\rho_r(\eta)},
\end{equation}
where $\eta$ represents the conformal time. Until the curvaton decays, it behaves as matter and $r$ increases proportionally to the scale factor. Therefore, the value of $r$ at the time of curvaton decay, $r_D\equiv r(\eta_{\mathrm{dec}})$, is calculated as
\begin{eqnarray}
	r_D&=&\frac{1}{6}\left(\frac{f\theta}{M_P}\right)^2\frac{T_R}{T_\mathrm{dec}}\quad\mathrm{for}\quad m_\sigma\geq\Gamma_I\label{rD1}\\
	r_D&=&\frac{1}{6}\left(\frac{f\theta}{M_P}\right)^2\frac{T_\mathrm{osc}}{T_\mathrm{dec}}\quad\mathrm{for}\quad m_\sigma\leq\Gamma_I,
\end{eqnarray}
where the subscripts ``$R$'', ``dec'' and ``osc'' represent the time of the reheating, the curvaton decay and the beginning of the curvaton oscillation. $\Gamma_I$ is the decay rate of the inflaton, and so $m_\sigma\gtrsim\Gamma_I$ ($m_\sigma\lesssim\Gamma_I$) stands for the case where the curvaton starts to oscillate before (after) the reheating. We used the reduced Planck mass $M_P=1/\sqrt{8\pi G}\simeq2.44\times10^{18}\,\mathrm{GeV}$. After the curvaton decays, $r$ becomes constant.
In this paper, we consider only the case $r_D<1$ for simplicity.

When the Hubble parameter becomes comparable to the curvaton decay rate $\Gamma_\sigma$, the curvaton decays into the radiation. We assume that the interaction of the curvaton is suppressed by $f$ like the axion and the decay rate is represented as
\begin{equation}
	\Gamma_\sigma=\frac{\kappa^2}{16\pi}\frac{m_\sigma^3}{f^2},\label{decay_rate}
\end{equation}
where $\kappa$ is a coupling constant less than unity (see Eqs.~\eqref{Lagrangian1}-\eqref{decay_rate_alpha} for the relation with the Lagrangian). The decay temperature is related to the decay rate as
\begin{equation}
	T_{\mathrm{dec}}=\left(\frac{90}{\pi^2}\right)^{\frac{1}{4}}g_*^{-\frac{1}{4}}(\Gamma_\sigma M_P)^{\frac{1}{2}}
	=\frac{90^\frac{1}{4}}{4\pi}g_*^{-\frac{1}{4}}\kappa\frac{M_P^\frac{1}{2}m_\sigma^{\frac{3}{2}}}{f}.
\end{equation}
where $g_*$ is the effective number of relativistic degrees of freedom.


\subsection{Curvature perturbation}\label{sec_curvature_perturbation}
In this model, the curvaton field dominantly contributes to the small-scale perturbations, which leads to PBH formation. On the other hand, the perturbations produced by the inflaton field are dominant at large scales and reproduce the CMB spectrum.
The curvaton has no coupling to the inflation.
The power spectrum of the curvature perturbation is given by the sum of the contributions from the curvaton and the inflaton:
\begin{equation}
	\mathcal P_\zeta(k)=\mathcal P_{\zeta, {\mathrm{inf}}}(k)+\mathcal P_{\zeta, {\mathrm{curv}}}(k).
\end{equation}
In the CMB observation scale, $\mathcal P_{\zeta, {\mathrm{inf}}}$ is almost scale-invariant and the amplitude is around $2\times10^{-9}$~\cite{Ade:2015xua}. 
We define $k_\mathrm{c}$ as the scale at which the contribution from the inflaton and that from the curvaton are comparable:
\begin{equation}
	\mathcal P_{\zeta, {\mathrm{curv}}}(k_\mathrm{c})\equiv\mathcal P_{\zeta, {\mathrm{inf}}}\simeq2\times10^{-9}.
\end{equation}
It is reasonable to require that the contribution from the inflaton is dominant at the scale larger than $1\,\mathrm{Mpc}$~\cite{nicholson2009reconstruction,nicholson2010reconstruction, bird2011minimally}:
\begin{equation}
	k_\mathrm{c}\gtrsim1\,\mathrm{Mpc^{-1}}.\label{1Mpc}
\end{equation}

\indent 
Let us discuss the generation of the curvature perturbation from the curvaton fluctuation. Since the energy density of the curvaton is $\rho_\sigma=m_\sigma^2\sigma^2/2$, its perturbation is expressed at linear order as
\begin{equation}
	\frac{\delta\rho_\sigma}{\rho_\sigma}=\frac{2\delta\sigma}{\sigma}=\frac{2\delta\theta}{\theta}.
\end{equation}
We consider the case where the density perturbation of the radiation is negligible compared with that of the curvaton. Then, the quantity above is nearly equal to the isocurvature perturbation induced by the curvaton,  $S$:
\begin{equation}
	S\equiv\frac{\delta\rho_\sigma}{\rho_\sigma}-\frac{3}{4}\frac{\delta\rho_r}{\rho_r}\simeq\frac{\delta\rho_\sigma}{\rho_\sigma}.\label{isocurvature}
\end{equation}
The power spectrum of $\delta \theta$ generated during the inflation is given as
\begin{equation}
	\mathcal P_{\delta\theta}^{1/2}(k)=\frac{H_\mathrm{inf}}{2\pi\varphi(k)},\label{Ptheta}
\end{equation}
where $\varphi(k)$ are evaluated at the time of the horizon exit of the scale $k$.
$\mathcal P_{\delta\theta}(k)$ is conserved on super-horizon scales. 
Then, the power spectrum of the isocurvature perturbations induced by the curvaton is
\begin{equation}
	\mathcal P_{S}^{1/2}(k)=\frac{H_\mathrm{inf}}{\pi\varphi(k)\theta}.
	\label{s_power_spec}
\end{equation}
As the curvature perturbation is expressed at linear order as
\begin{equation}
	\zeta=-H\frac{\delta\rho}{\dot\rho},\label{zeta_flat}
\end{equation}
where the right hand side is evaluated on the flat slicing of the spacetime, it is rewritten by using $\rho=\rho_r+\rho_\sigma$ and the Friedmann equations as
\begin{equation}
	\zeta=\frac{\delta\rho_r+\delta\rho_\sigma}{4\rho_r+3\rho_\sigma}\simeq\frac{\delta\rho_\sigma}{4\rho_r+3\rho_\sigma}.
\end{equation}
Then, the power spectrum of the curvature perturbations contributed by the curvaton is obtained as
\begin{equation}
	\mathcal P_{\zeta,\mathrm{curv}}(k,\eta)=\left(\frac{r(\eta)}{4+3r(\eta)}\right)^2\mathcal P_{S}(k)=\left(\frac{2r(\eta)}{4+3r(\eta)}\right)^2\left(\frac{H_\mathrm{inf}}{2\pi\varphi(k)\theta}\right)^2,
	\label{Pzeta_eta}
\end{equation}
where in the radiation-dominated era, which follows the reheating phase, $r(\eta)$ behaves as
\begin{align}
r(\eta) =
\begin{cases}
\displaystyle
r_D \frac{\eta}{\eta_\text{dec}}
&\text{for}\,\,\,\eta \leq \eta_\text{dec},\\
\displaystyle
r_D
&\text{for}\,\,\,\eta> \eta_\text{dec}.\\
\end{cases}
\label{r_evol}
\end{align}
Note that the curvature perturbation is not conserved even on super-horizon scales because the fraction of the energy density of the curvaton increases in the radiation before the decay of the curvaton. 
After the curvaton decays into the radiation, the spectrum becomes constant in time and we denote the value at that time just removing the argument $\eta$ as
\begin{equation}
	\mathcal P_{\zeta,\mathrm{curv}}(k)\equiv\mathcal P_{\zeta,\mathrm{curv}}(k,\eta>\eta_\mathrm{dec})
	=\left(\frac{r_D}{4+3r_D}\right)^2\mathcal P_{S}(k)
	=\left(\frac{2r_D}{4+3r_D}\right)^2\left(\frac{H_\mathrm{inf}}{2\pi\varphi(k)\theta}\right)^2.\label{Pzeta}
\end{equation}
 After the field $\varphi$ reaches its minimum $\varphi_\mathrm{min}\simeq f$ during inflation, the value of $\varphi(k)$ in Eq.~\eqref{Ptheta} becomes constant. Therefore, the spectrum is given by
 \begin{equation}
 	\mathcal P_{\zeta,\mathrm{curv}}(k)=\mathcal P_{\zeta,\mathrm{curv}}(k_*)=\left(\frac{2r_D}{4+3r_D}\right)^2\left(\frac{H_\mathrm{inf}}{2\pi f\theta}\right)^2\quad \mathrm{for}\quad k>k_*,\label{Pzeta_flat}
\end{equation}
where $k_*$ indicates the scale which exits the horizon when $\varphi$ reaches its minimum. For the scale $k<k_*$, the power spectrum has scale dependence. We denote it using the spectral index $n_\sigma$ as
\begin{equation}
	\mathcal P_{\zeta,\mathrm{curv}}(k)=\mathcal P_{\zeta,\mathrm{curv}}(k_c)\left(\frac{k}{k_c}\right)^{n_\sigma-1}\quad \mathrm{for}\quad k<k_*.\label{Pzeta_blue}
\end{equation}
Here we remark that Eqs.~\eqref{Pzeta_flat} and \eqref{Pzeta_blue} are actually valid only for super-horizon modes at each time of $\eta>\eta_\text{dec}$ because $\zeta$ is not a conserved quantity on sub-horizon scales.

\begin{figure}
	\centering
	\includegraphics[width=.45\textwidth]{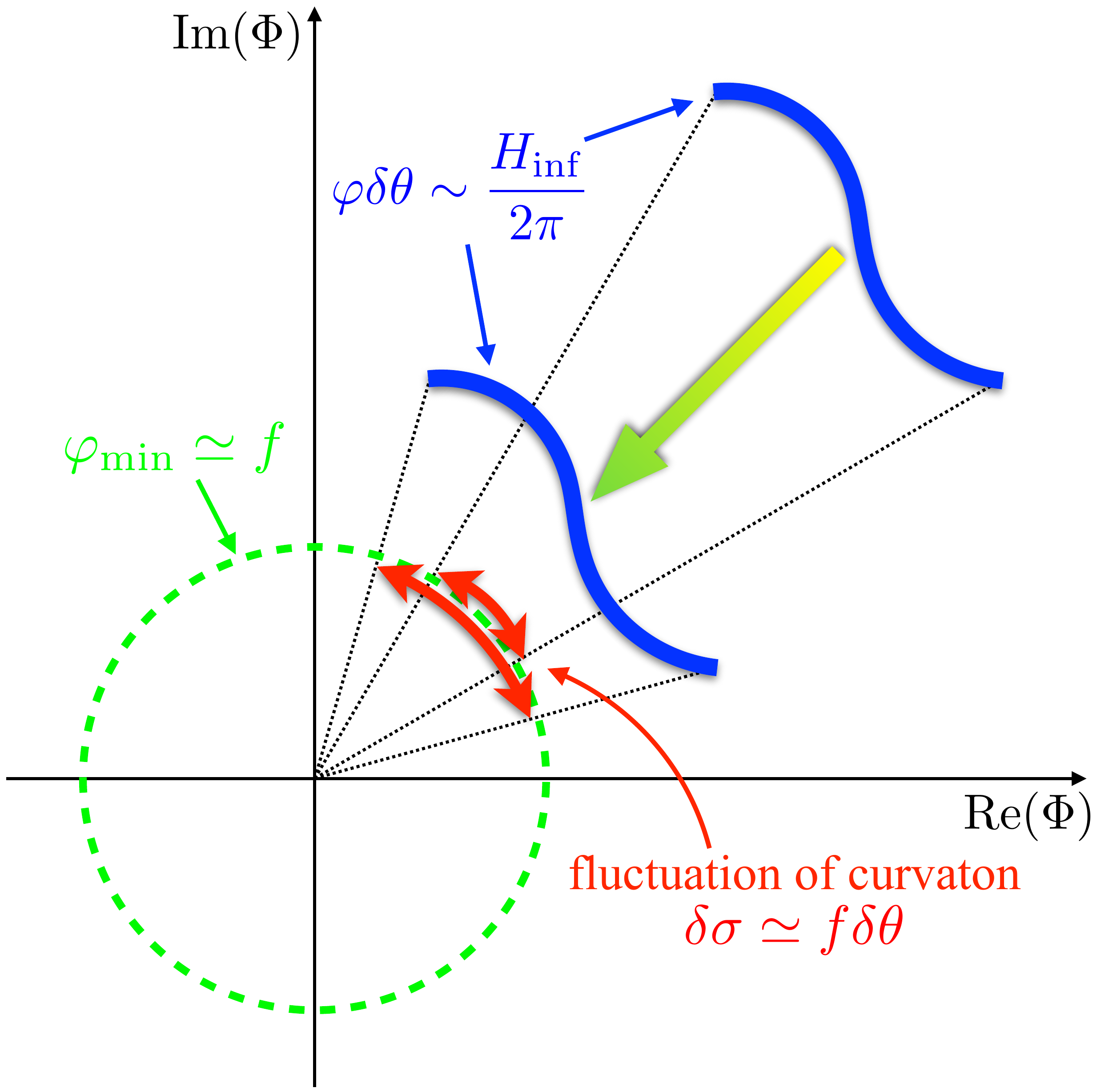}
	\caption{\small
	A schematic diagram showing the mechanism for a blue-tilted spectrum in the axion-like curvaton model.
	Fluctuation of curvaton $\sigma$, corresponding to the angular direction, is indicated by red thick arrows.
	During inflation, the radial field $\varphi$ rolls down the potential given in Eq.~\eqref{SUGRA} from a large value to $\varphi_\mathrm{min}\simeq f$.
	We can see from this figure that the motion of $\varphi$ makes $\delta\sigma$ larger and larger (see Eq.~\eqref{Ptheta}).
	Also, note that smaller-scale perturbations exit the horizon at a later time during inflation.
	As a result, smaller-scale perturbations come to obtain larger amplitude.
	}
	\label{fig:Axion-curvaton}
\end{figure}

The spectral index can be calculated solving the equation of motion of $\varphi$. From the potential Eq.~\eqref{SUGRA}, the equation of motion is written as
\begin{equation}
	\ddot\varphi+3H\dot\varphi+cH^2\varphi=0.
\end{equation}
Taking account of $H\simeq \mathrm{const.}$ during the inflation and $k=aH\propto e^{Ht}$ at the horizon exit, the solution is obtained as
\begin{equation}
	\varphi\propto e^{-\lambda Ht}\propto k^{-\lambda} \quad\mathrm{with}\quad \lambda=\frac{3}{2}-\frac{3}{2}\sqrt{1-\frac{4}{9}c}.\label{phi_solution}
\end{equation}
Combining Eqs.~\eqref{Pzeta}, \eqref{Pzeta_blue} and \eqref{phi_solution}, the spectral index $n_\sigma$ is obtained as
\begin{equation}
	n_\sigma-1=3-3\sqrt{1-\frac{4}{9}c},\label{tilt}
\end{equation}
which shows that a blue-tilted spectrum of the curvature perturbations such as $n_\sigma\simeq2-4$ is realized for $c$ of the order of unity.
In Fig.~\ref{fig:Axion-curvaton}, the mechanism for a blue-tilted spectrum explained above is displayed schematically.  

\begin{figure}
	\centering
	\includegraphics[width=.45\textwidth]{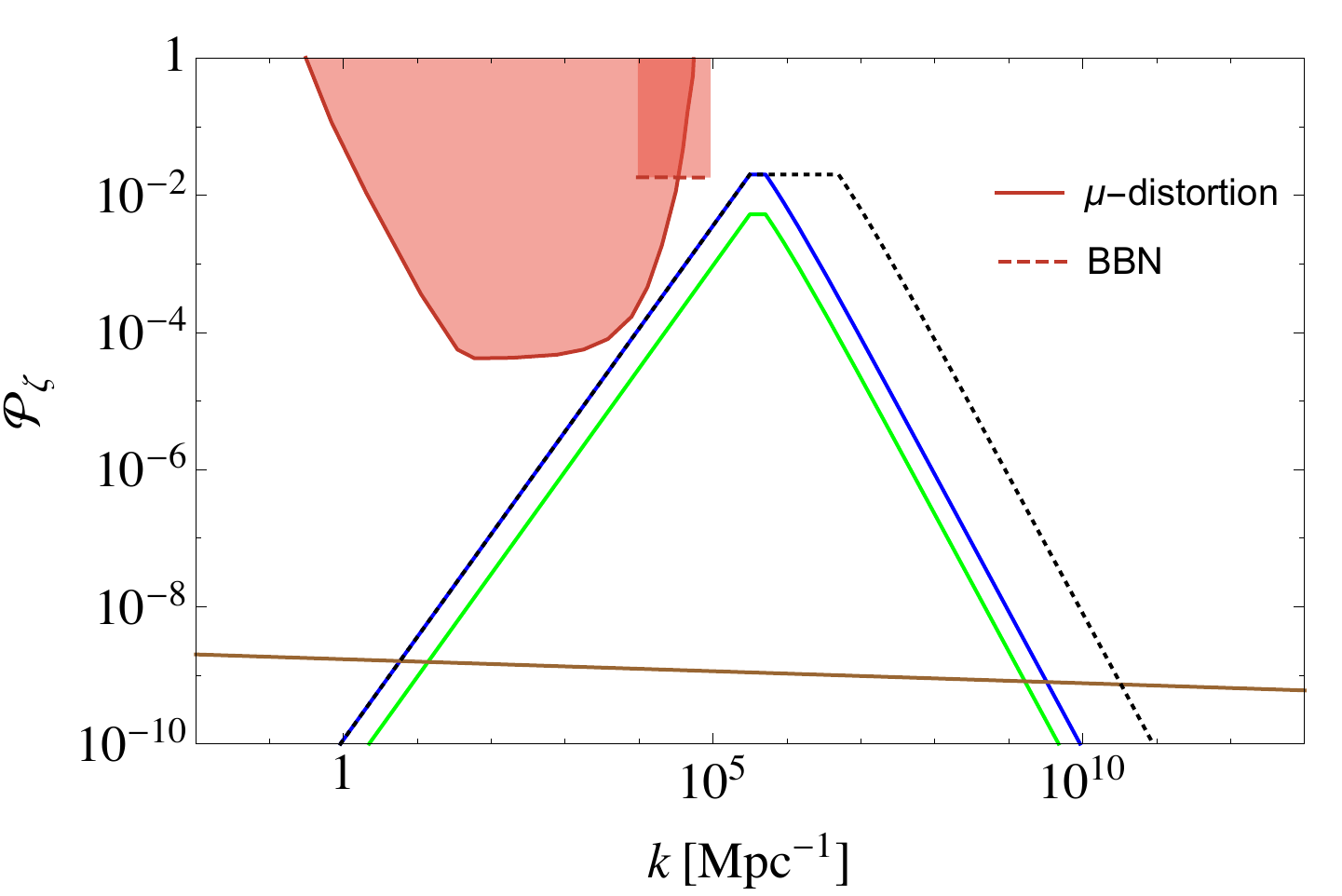}
	\caption{\small
	{\bf Green solid line:} the curvature perturbation spectrum induced by the curvaton at the horizon reentering of each mode, $\mathcal P_{\zeta,\mathrm{curv}}(k,\eta=k^{-1})$,  for parameters given in Eq.~\eqref{parameter}. 
	In this line, non-Gaussianity is taken into account following the procedure given in Sec.~\ref{subsec:NG_suppression}.
	{\bf Blue solid line:} the same as the green solid line with the amplitude changed so that the same abundance of  PBHs is realized when we assume the Gaussian distribution (see also the discussion given below Eq.\eqref{Q}).
	 {\bf Black dotted line:} the same as the blue solid line with $k_\mathrm{dec}$ replaced by $5\times 10^{6}\,\mathrm{Mpc^{-1}}$. 
	 {\bf Brown solid line:} the curvature perturbation spectrum induced by the inflaton. 
	 {\bf Red shaded regions:} the constraints from the CMB $\mu\mathchar`-$distortion~\cite{Kohri:2014lza,Fixsen:1996nj}
	  and BBN~\cite{Inomata:2016uip} (see also~\cite{Jeong:2014gna, Nakama:2014vla}).
	}
	\label{fig:curvature_spectrum2}
\end{figure}

The curvature perturbation spectrum at the horizon reentering of each mode, $\mathcal P_{\zeta,\mathrm{curv}}(k,\eta=k^{-1})$, is shown in Fig.~\ref{fig:curvature_spectrum2} for parameters given in Eq.~\eqref{parameter}.
Note that PBH production rate is determined by the amplitude of the perturbations at the horizon reentering 
and therefore $\mathcal P_{\zeta,\mathrm{curv}}(k,\eta=k^{-1})$ dominantly determines the PBH abundance of each mass.
In general, the shape of the spectrum becomes a plateau like the black dotted line in Fig.~\ref{fig:curvature_spectrum2} because the spectrum is flat for the region between $k_*$ and $k_\mathrm{dec}\equiv\eta_\mathrm{dec}^{-1}$.
As will be mentioned in Sec~\ref{sec_Result}, for a realistic scenario to reproduce the LIGO events, the flat region is required to be much narrower like the blue solid line in Fig.~\ref{fig:curvature_spectrum2} in order to avoid the constraints from the CMB $\mu$-distortion, BBN, and the pulsar timing array (PTA) experiments (see also Fig.~\ref{fig:GW_spectrum}). 
Moreover, as will be mentioned in Sec~\ref{subsec:NG_suppression} and Sec~\ref{sec_Result}, $\mathcal P_\zeta$ is suppressed for the fixed abundance of PBH if we take account of non-Gaussianity. 
Then, the green solid line is physically realized.

\subsection{Non-Gaussianity}
\label{subsec:non-Gaussianity}

It is known that curvatons tend to generate somewhat large non-Gaussianity.  
We are now assuming the quadratic potential of the curvaton as Eq.~\eqref{curvaton_potential}. 
The dominant contribution of non-Gaussianity comes only from the quadratic one\footnote{
This can be understood from $\frac{\delta\rho_\sigma}{\rho_\sigma}=\frac{2\delta\sigma}{\sigma}+\left(\frac{\delta\sigma}{\sigma}\right)^2$ at full order.
 However, it is not appropriate to insert this expression into Eq.~\eqref{zeta_flat} because that equation is valid only at linear order. 
 Therefore, we use the $\delta N$ formalism.
}
 such as
\begin{equation}
	\zeta(\bvec x)=\zeta_g(\bvec x)+\frac{3}{5}f_\mathrm{NL}\left(\zeta_g^2(\bvec x)-\left<\zeta_g^2(\bvec x)\right>\right),\label{quadratic_NG1}
\end{equation}
where $\zeta_g(\bvec x)$ has a Gaussian distribution and $\left<\zeta_g^2(\bvec x)\right>$ is subtracted to ensure $\left<\zeta_g(\bvec x)\right>=0$. In this section, we review how $f_\mathrm{NL}$ is estimated and the discussion is also valid in other models as far as the potential is quadratic.

The curvature perturbation is calculated using the $\delta N$ formalism as
\begin{equation}
	\zeta=\delta N=\frac{\partial N}{\partial \sigma}\delta \sigma+\frac{1}{2}\frac{\partial^2 N}{\partial \sigma^2}\left(\delta\sigma^2-\left<\delta\sigma^2\right>\right),\label{quadratic_deltaN}
\end{equation}
where $\delta N$ means the fluctuation of the number of e-folds from an initial flat slice to a final uniform density slice, and $\delta\sigma$ is the fluctuation of the curvaton at the onset of the oscillation. 
Since $\delta\sigma$ follows a Gaussian distribution, comparing Eqs.~\eqref{quadratic_NG1} and \eqref{quadratic_deltaN}, $f_\mathrm{NL}$ is expressed as
\begin{equation}
	f_\mathrm{NL}=\frac{5}{6}\frac{\partial^2 N}{\partial \sigma^2}\left(\frac{\partial N}{\partial \sigma}\right)^{-2}.\label{fNL_deltaN}
\end{equation}

As mentioned earlier, we denote the time of the onset of the curvaton oscillation as $t_\mathrm{osc}$ and the time of the curvaton decay as $t_\mathrm{dec}$.\footnote{
Since the produced PBH abundance is dominantly determined by the modes entering horizon after the curvaton decay,
 we focus on the modes with $k<k_\text{dec}$ in the context of non-Gaussianity.
}
 The Friedmann equation $3M_P^2H^2=\rho$ is rewritten at $t=t_\mathrm{osc}$ and $t=t_\mathrm{dec}$ as
\begin{eqnarray}
	3m_\sigma^2M_P^2&=&\rho_{r,\mathrm{osc}}+\rho_{\sigma,\mathrm{osc}}\simeq\rho_{r,\mathrm{osc}}\label{Friedmann_t1}\\
	3\Gamma_\sigma^2M_P^2&=&\rho_{r,\mathrm{dec}}+\rho_{\sigma,\mathrm{dec}}=\left(\frac{a_\mathrm{osc}}{a_\mathrm{dec}}\right)^4\rho_{r,\mathrm{osc}}+\left(\frac{a_\mathrm{osc}}{a_\mathrm{dec}}\right)^3\rho_{\sigma,\mathrm{osc}},\label{Friedmann_t2}
\end{eqnarray}
where the subscripts ``$\mathrm{osc}$'' and ``$\mathrm{dec}$'' stand for the value at the time of $t=t_\mathrm{osc}$ and $t=t_\mathrm{dec}$. As mentioned above, the ratio $r\equiv\rho_\sigma/\rho_r$ increases in proportion to the scale factor. Then, the energy density of the radiation, which is almost homogeneous, dominates over that of the curvaton at $t=t_\mathrm{osc}$, and the uniform density slice determined by the constant Hubble parameter $H=m_\sigma$ is also the spatially flat slice, which is required for the $\delta N$ formula. 
We define the number of e-folds $N$ between $t=t_\mathrm{osc}$ and $t=t_\mathrm{dec}$ as
\begin{equation}
	e^N\equiv\frac{a_\mathrm{dec}}{a_\mathrm{osc}}.
\end{equation}
From Eq.~\eqref{Friedmann_t2}, it is noted that
\begin{equation}
	e^{-4N}+e^{-3N}\frac{\rho_{\sigma,\mathrm{osc}}}{\rho_{r,\mathrm{osc}}}=\mathrm{const.}\label{independent_sigma}
\end{equation}
is independent of $\sigma$. $\rho_{r,\mathrm{osc}}$ is independent of $\sigma$ at leading order because of Eq.~\eqref{Friedmann_t1}. Taking account of $\rho_{\sigma}\propto\sigma^2$, one can differentiate Eq.~\eqref{independent_sigma} with respect to $\sigma$ and obtain
\begin{eqnarray}
	\frac{\partial N}{\partial\sigma}&=&\frac{2}{\sigma}\frac{r_D}{4+3r_D}\\
	\frac{\partial^2 N}{\partial\sigma^2}&=&\frac{2r_D}{\sigma^2}\left[\frac{8r_D}{(4+3r_D)^3}-\frac{6r_D}{(4+3r_D)^2}+\frac{1}{4+3r_D}\right],
\end{eqnarray}
where we use $r_D=e^N\rho_{\sigma,\mathrm{osc}}/\rho_{r,\mathrm{osc}}$. Then, with the help of Eq.~\eqref{fNL_deltaN}, one obtains 
\begin{equation}
	f_{\mathrm{NL}}=\frac{5}{12}\left(-3+\frac{4}{r_D}+\frac{8}{4+3r_D}\right)\label{NG}.
\end{equation}


\section{PBH formation}\label{PBH_formation}
PBHs are formed if the density perturbation averaged over a Hubble patch is larger than a threshold value $\delta_c$ when the overdense region reenters the horizon. The mass of PBH is proportional to the horizon mass at the formation:
\begin{eqnarray}
	M&=&\gamma\frac{4\pi}{3}\rho_r H^{-3}\nonumber\\
	&\simeq&6\times10^{35} \,\mathrm g\left(\frac{\gamma}{0.2}\right)\left(\frac{g_*}{10.75}\right)^{-\frac{1}{2}}\left(\frac{T}{10^{-2}\,\mathrm{GeV}}\right)^{-2}\nonumber\\
	&\simeq&8\times10^{35} \,\mathrm g\left(\frac{\gamma}{0.2}\right)\left(\frac{g_*}{10.75}\right)^{-\frac{1}{6}}\left(\frac{k}{10^5\,\mathrm{Mpc^{-1}}}\right)^{-2}\nonumber\\
	&\simeq&8\times10^{35} \,\mathrm g\left(\frac{\gamma}{0.2}\right)\left(\frac{t}{0.01\,\mathrm{s}}\right),\label{PBHmass}
\end{eqnarray}
where $\gamma$ is the proportionality constant. 
The values of $\delta_c$ and $\gamma$ have been investigated and have some uncertainties. 
In this paper, we use the following values $\delta_c=0.4$~\cite{Harada:2013epa} and $\gamma=3^{-3/2}\simeq0.2$~\cite{Carr:1975qj}.

The coarse-grained density perturbation smoothed over the scale $R$ is defined as
\begin{eqnarray}
	\delta_\mathrm{cg}(\bvec x;R)&\equiv&\int\mathrm d^3x'W(|\bvec x-\bvec x'|;R)\delta(\bvec x')\\
	&=&\int\frac{\mathrm d^3k}{(2\pi)^3}\tilde W(k;R)\tilde\delta(\bvec k)e^{i\bvec k\cdot \bvec x},
\end{eqnarray}
where $W(x;R)$ is a window function and $\tilde W(k;R)$ is its Fourier mode.
The correlation function is given by
\begin{equation}
	\left<\delta_\mathrm{cg}(\bvec x_1;R)\delta_\mathrm{cg}(\bvec x_2;R)\right>=\int\frac{\mathrm dk}{k}\tilde W^2(k;R)\mathcal P_\delta(k)\frac{\sin(k|\bvec x_1-\bvec x_2|)}{k|\bvec x_1-\bvec x_2|}.
\end{equation}

The abundance of PBHs is characterized by $\beta$, which is the production rate of PBHs at the time of the formation. Assuming a Gaussian probability density function (PDF), $\beta$ is calculated according to the Press-Schechter theory~\cite{Press:1973iz} as
\begin{eqnarray}
	\beta(M)
	&=&\int_{\delta_c}\mathrm d\delta\frac{1}{\sqrt{2\pi\left<\delta_{\mathrm{cg}}^2(M)\right>}}\exp\left(-\frac{\delta^2}{2\left<\delta_{\mathrm{cg}}^2(M)\right>}\right)\nonumber\\
	&\simeq&\frac{1}{\sqrt{2\pi}}\frac{\sqrt{\left<\delta_{\mathrm{cg}}^2(M)\right>}}{\delta_c}\exp\left(-\frac{\delta_c^2}{2\left<\delta_{\mathrm{cg}}^2(M)\right>}\right).\label{beta}
\end{eqnarray}
Here, $\left<\delta_{\mathrm{cg}}^2(M)\right>$ is the variance of the coarse-grained density perturbation smoothed over the horizon scale\footnote{\label{delta_not_zeta}
Two different quantities, the curvature perturbation $\mathcal R$ and the density perturbation $\delta$ both in the comoving gauge, have been used in order to determine the criterion for PBH formation.
 From the Einstein equation, they are related through $\delta=\frac{4}{9}\left(\frac{k}{aH}\right)^2\mathcal R$ on super-horizon scales. 
 The comoving curvature perturbation $\mathcal R$ coincides with the adiabatic curvature perturbation $\zeta$ on super-horizon scales.
  When $\mathcal R$ is smoothed, the factor $(qR)^4$ of Eq.~\eqref{delta_cg1} gets outside of the integral and becomes $(kR)^4=1$. 
  This is the case in Eq.~(34) in Ref.~\cite{Kawasaki:2012wr}. 
  However, it is recently suggested that it should be more appropriate to use $\delta$~\cite{Young:2014ana}. 
  Then, in this paper, we follow the convention of smoothing $\delta$, which leads to $(qR)^4$ inside the integral in Eq.~\eqref{delta_cg1}.}
and is written as
\begin{align}
	\left<\delta_{\mathrm{cg}}^2(M)\right>=\int_0^\infty \frac{\mathrm dq}{q}\,\tilde W^2(q;R)\frac{16}{81}(qR)^4\mathcal P_\zeta(q),\label{delta_cg1}
\end{align}
where 
$R=k^{-1}=(aH)^{-1}$ is the horizon scale associated with the PBH mass $M$ through Eq.~\eqref{PBHmass}. 
One choice of a window function\footnote{
Uncertainties on choice of a window function is investigated in detail in Ref.~\cite{Ando:2018qdb}.
}
is the Gaussian window function $W(x;R)=\exp(-x^2/(2R^2))/(\sqrt{2\pi}R)^3$, 
which corresponds to $\tilde W(k;R)=\exp(-k^2R^2/2)$.
In the axion-like curvaton model, $\left<\delta_{\mathrm{cg}}^2\right>$ for the Gaussian window function is analytically calculated using Eqs.~\eqref{Pzeta_flat} and \eqref{Pzeta_blue} as 
\begin{equation}
	\left<\delta_{\mathrm{cg}}^2(R)\right>=\frac{8}{81} \left(\frac{r(R)}{4+3r(R)}\right)^2 \mathcal{P}_S(k_*)\left[(k_*R)^{-(n_\sigma-1)}\gamma\left(\frac{n_\sigma+3}{2},k_*^2R^2\right)+\hat\gamma(2,k_*^2R^2)\right]\label{improved}
\end{equation}
where
\begin{equation}
	\gamma(a,x)=\int_0^xt^{a-1}e^{-t}\mathrm dt\,,\quad\hat\gamma(a,x)=\int_x^\infty t^{a-1}e^{-t}\mathrm dt.
	\label{eq:def_gamma_funcs}
\end{equation}
Another choice of a window function is a top-hat one $W(x;R)=\frac{3}{4\pi R^3}\Theta(R-x)$, which corresponds to $\tilde W(k;R)=\frac{3}{(kR)^3}(\sin (kR)-kR\cos(kR))$. Here, $\Theta(x)$ is the step function.

\begin{figure}
	\centering
	\includegraphics[width=.45\textwidth]{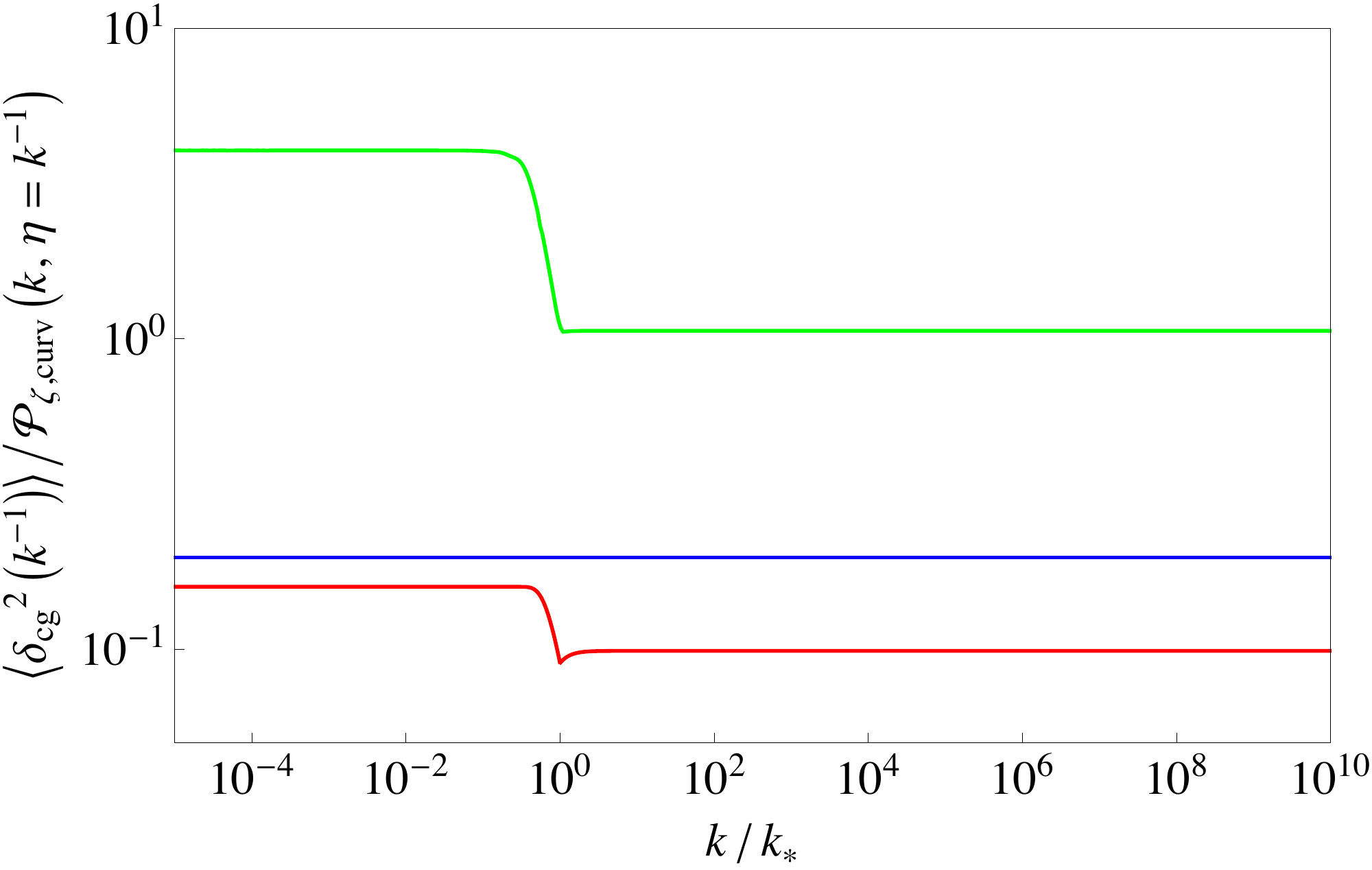}
	\caption{\small
	The $k/k_*$ dependence of $\vev{\delta_\text{cg}^2(k^{-1})}/ \mathcal P_{\zeta,\text{curv}}(k,\eta=k^{-1})$ in the case of $n_\sigma=2.5$. 
	{\bf Red solid line:} for the Gaussian window function. 
	This is calculated by using the analytical expression, Eq.~\eqref{improved}.
	{\bf Green solid line:} for the top-hat window function.
	This is calculated by applying the top-hat window function to Eq.~\eqref{delta_cg1} and also tracing the sub-horizon time evolution of perturbations.
	{\bf Blue solid line:} for the rough estimate of the relation given by Eq.~\eqref{16/81}.
	}
	\label{fig:coarse_grained}
\end{figure}
Fig.~\ref{fig:coarse_grained} shows the $k/k_*$ dependence of $\vev{\delta_\text{cg}^2(k^{-1})}/ \mathcal P_{\zeta,\text{curv}}(k,\eta=k^{-1})$ in the case of $n_\sigma=2.5$. 
The red solid line is for the Gaussian window function.
From this, we can see that the approximate relation $\left<\delta_{\mathrm{cg}}^2\right>\simeq0.1\,\mathcal P_{\zeta,\mathrm{curv}}$ holds almost independently of $k$.\footnote{
There is no increase of the ratio of $\left<\delta_{\mathrm{cg}}^2(k^{-1})\right>$ to $\mathcal P_{\zeta,\mathrm{curv}}(k)$ on small scales, which is seen in Fig. 1 of Ref.~\cite{Kawasaki:2012wr}.
}
This result is different from Fig.~1 (and Eq.~(34)) in Ref.~\cite{Kawasaki:2012wr} because the coarse-grained curvature perturbations  was used to estimate $\langle \delta^2_\text{cg}\rangle$ in Ref.~\cite{Kawasaki:2012wr}.
In contrast, we have directly obtained the coarse-grained density perturbation through the relation $\delta=\frac{4}{9}\left(\frac{k}{aH}\right)^2\mathcal R$.
Meanwhile, the green solid line is for the top-hat window function. 
Since the density perturbations coarse-grained by the top-hat window function is sensitive to scales well within the horizon, we trace the sub-horizon time evolution of perturbations by multiplying the integrand of Eq.~\eqref{delta_cg1} by the convolution of the transfer function.\footnote{
As discussed in Sec.~\ref{sec_GW}, the transfer function in the curvaton model is complicated. However, here we just use the standard transfer function in the radiation-dominated universe for simplicity: $T(k,\eta)=\frac{3}{X^3}(\sin X-X\cos X)$ where $X=k\eta/\sqrt 3$.
}
In this case, the relation $\left<\delta_{\mathrm{cg}}^2\right>\simeq1\times\mathcal P_{\zeta,\mathrm{curv}}$ holds for the scale where the curvature perturbation is scale-invariant. 
The factor is larger than that in the case of the Gaussian window function and agrees with the result in Ref.~\cite{Blais:2002gw}.
In Fourier space, the top-hat window function decays slowly on small scales, compared to the Gaussian window function.
Hence, the ratio of $\left<\delta_{\mathrm{cg}}^2\right>$ to $\mathcal P_{\zeta,\mathrm{curv}}$ calculated with the top-hat window function is larger than 
that calculated with the Gaussian window function.
In the following, since it is not obvious which window function should be used, 
we adopt the following simple relation:
\begin{equation}
	\vev{\delta_\text{cg}^2(k^{-1})}\simeq\frac{16}{81}\mathcal P_{\zeta,\text{curv}}(k,\eta=k^{-1}).\label{16/81}
\end{equation}
where this expression is derived from the rough estimate of the relation between the density perturbations and curvature perturbations at the horizon crossing $\delta \simeq \frac{4}{9} \mathcal \zeta$.
We represent the relation given by Eq.\,\eqref{16/81} with a blue solid line in Fig.~\ref{fig:coarse_grained}.
It can be seen that Eq.\,\eqref{16/81} corresponds to the intermediate case between the Gaussian window function and the top-hat window function.

Considering that PBHs behave as matter, one can evaluate the current abundance of PBHs with mass $M$ over the logarithmic interval $\ln M$ as
\begin{align}
	\frac{\Omega_\mathrm{PBH}(M)}{\Omega_c}
	&\simeq \left. \frac{\rho_\text{PBH}}{\rho_m} \right|_\text{eq} \frac{\Omega_m}{\Omega_c} =  \left( \frac{T_M}{T_\text{eq}} \frac{\Omega_m}{\Omega_c} \right) \gamma \beta(M)  \nonumber \\
	&\simeq \left(\frac{\beta(M)}{1.84\times10^{-8}}\right)\left(\frac{\gamma}{0.2}\right)^{\frac{3}{2}}\left(\frac{10.75}{g_*}\right)^{\frac{1}{4}}
	\left(\frac{0.12}{\Omega_ch^2}\right)\left(\frac{M}{M_\odot}\right)^{-\frac{1}{2}},
	\label{Omega_PBH}
\end{align}
where $\Omega_\mathrm{PBH}\,, \Omega_c $ and $\Omega_m$ are the current density parameters of PBHs, DM and total matter (DM+baryon) respectively,
and $T_M$ and $T_\text{eq}$ are the temperatures at the PBH formation with $M$ and the matter-radiation equality time.
Note that $\gamma$ appearing in Eq.~(\ref{Omega_PBH}) is different from $\gamma(a,x)$ which is defined in Eq.~(\ref{eq:def_gamma_funcs}).
 The total abundance of PBHs is expressed as
\begin{equation}
	\Omega_{\mathrm{PBH,\,tot}}=\int\mathrm d\ln M\Omega_\mathrm{PBH}(M).
\end{equation}

\section{Gravitational waves from second-order effects}\label{sec_GW}

If there are large scalar perturbations that form PBHs, their second-order effect can be a source of primordial GWs. 
In this section, the GWs generated in the axion-like curvaton model are investigated quantitatively mainly following Ref.~\cite{Kawasaki:2013xsa}. 
Most parts of the discussion can be seen in the literature such as Ref.~\cite{Ananda:2006af,Baumann:2007zm,Saito:2008jc,Saito:2009jt,Bugaev:2009zh,Bugaev:2010bb,Inomata:2016rbd} except that the origin of primordial perturbations is isocurvature perturbations in the curvaton model.
 Then, we review the general argument about GWs induced by scalar perturbations in App.~\ref{app_GW} and describe characteristic features of the curvaton model here.

We define a transfer function $T(\eta,k)$ as\footnote{\label{magnitude_k}
Since we consider the homogeneous and isotropic universe, the transfer function $T$ depends on only $\eta$ and the magnitude of $k$.
}
\begin{equation}
	\Psi(\eta,\bvec k)=T(\eta,k)S(\bvec k),\label{trans_Z}
\end{equation}
where $\Psi$ is the curvature perturbation in the conformal Newtonian gauge and $S$ is the isocurvature perturbation induced by the curvaton defined by Eq.~\eqref{isocurvature}.
The isocurvature perturbation is almost conserved in the case of $r_D<1$ \cite{dodelson}.
By using the transfer function, we can express the power spectrum of GWs as (see Eq.~\eqref{Green}, \eqref{f} and \eqref{P_h})
\begin{eqnarray}
	\mathcal P_h(\eta,k)&=&\frac{1}{4}\int_0^\infty dy\int_{|1-y|}^{1+y}dx\frac{y^2}{x^2}\left(1-\frac{(1+y^2-x^2)^2}{4y^2}\right)^2\mathcal P_S(kx)\mathcal P_S(ky)
	\left[\frac{k^2}{a(\eta)}\int^\eta d\tilde\eta a(\bar\eta)g_{\bvec k}(\eta;\bar\eta)f(ky,kx,\bar\eta)\right]^2 \nonumber\\
	&=&\frac{1}{4}\int_0^\infty dy\int_{|1-y|}^{1+y}dx\frac{y^2}{x^2}\left(1-\frac{(1+y^2-x^2)^2}{4y^2}\right)^2
	\left(\frac{4+3r_D}{r_D}\right)^4\mathcal P_{\zeta,\mathrm{curv}}(kx)\mathcal P_{\zeta,\mathrm{curv}}(ky)\nonumber\\
	&&\hspace{80mm}\times\left[\frac{k^2}{a(\eta)}\int^\eta d\tilde\eta a(\bar\eta)g_{\bvec k}(\eta;\bar\eta)f(ky,kx,\bar\eta)\right]^2,
	\label{P_h_Z}	
\end{eqnarray}
where
\begin{equation}
	g_{\bvec k}(\eta;\bar\eta)\equiv\frac{\sin\left[k(\eta-\bar\eta)\right]}{k}\theta(\eta-\bar\eta)\label{Green_Z}
\end{equation}
and
\begin{equation}
	f(k_1,k_2,\eta)\equiv 4\left[3T(k_1)T(k_2)
	+\frac{2}{\mathcal H}T'(k_1)T(k_2)+\frac{1}{\mathcal H^2}T'(k_1)T'(k_2)\right].\label{f_Z}
\end{equation}
Here, $\eta$ dependence of $T(\eta,k)$ is made implicit.

As shown in Eq.~\eqref{Omega_GW}, the current density parameter of the GWs is written in terms of the power spectrum as
\begin{equation}
	 \Omega_{\mathrm{GW}}(\eta_0,k)
	 =0.83\prn{ \frac{g_{*,\star}}{10.75}    }^{-1/3}\frac{\Omega_{r,0}k^2}{24\mathcal H(\eta_\star)^2}\overline{\mathcal P_h(\eta_\star,k)},\label{Omega_GW_Z}
\end{equation}
where the overline stands for the time average and $\eta_\star$ is a certain time soon after the GWs begin to behave as radiation, $\Omega_\text{GW} \propto a^{-4}$.

\subsection{Transfer Function}

Let us discuss the transfer function emerging in the axion-like curvaton model. The transfer function $T(\eta,k)$ has different forms before and after the curvaton decay, which we denote as
\begin{equation}
	T(\eta,k)=
		\begin{cases}
			T_S(\eta,k)\quad\mathrm{for}\quad\eta<\eta_\mathrm{dec}\\
			T_\Psi(\eta,k)\quad\mathrm{for}\quad\eta>\eta_\mathrm{dec}
		\end{cases}.
\end{equation}

After the curvaton decays ($\eta>\eta_\mathrm{dec}$), the general solution for $\Psi$ is a linear combination of two modes, and so $T_\Psi$ can be written as
\begin{equation}
	T_\Psi(\eta,k)=C_1(k)\frac{\sin X-X\cos X}{X^3}+C_2(k)\frac{\cos X+X\sin X}{X^3}\quad\mathrm{with}\quad X\equiv\frac{k\eta}{\sqrt3}.\label{T_Psi}
\end{equation}
The coefficients $C_1(k)$ and $C_2(k)$ are determined by the continuity of $\Psi$ and its time derivative, i.e. $T_S=T_\Psi$ and ${T_S}'={T_\Psi}'$ at $\eta=\eta_\mathrm{dec}$.

The derivation of $T_S(\eta,k)$ is more complicated. It is evaluated by interpolating the super-horizon and sub-horizon solutions~\cite{Kawasaki:2013xsa}.
First, we will derive the super-horizon ($k\eta\ll1$) solution. The adiabatic curvature perturbation $\zeta$ and the comoving curvature perturbation $\mathcal R$ are written in the gauge invariant ways as
\begin{eqnarray}
	\zeta&=&-\Psi-\mathcal H\frac{\delta\rho}{\rho'}\\
	\mathcal R&=&-\Psi+\mathcal H\frac{\delta u}{a},
\end{eqnarray}
where $\delta u$ is the perturbation of the four-velocity called the velocity potential defined by
\begin{equation}
	u_\mu=(-a+a\delta u_0\,,\,\partial_i\delta u+\delta u_{i})\quad\mathrm{with}\quad\partial_i\delta u_i=0.
\end{equation}
From the Einstein equation, $\delta u$ is related to $\Psi$ as~\cite{mukhanov}
\begin{equation}
	\delta u=-\frac{2M_P^2(a\Psi)'}{a^2(\rho+p)}.
\end{equation}
In the axion-like curvaton model, this leads to
\begin{equation}
	\delta u=-\frac{2+2r}{4+3r}\frac{(a\Psi)'}{\mathcal H^2}.
\end{equation}
Then, the comoving curvature perturbation is expressed as
\begin{equation}
	\mathcal R=-\frac{6+5r}{4+3r}\Psi-\frac{2+2r}{4+3r}\frac{\Psi'}{\mathcal H}.\label{RR}
\end{equation}
On the other hand, as shown in Sec.~\ref{sec_curvature_perturbation}, the adiabatic perturbation is expressed as
\begin{equation}
	\zeta=\zeta_\mathrm{inf}+\frac{r}{4+3r}S\label{zetazeta}.
\end{equation}
One can show that
\begin{equation}
	\zeta-\mathcal R=-\mathcal H\frac{\delta\rho_\mathrm{com}}{\rho'}=\frac{2M_P^2\Delta\Psi}{3(\rho+p)a^2}\sim\left(\frac{k}{\mathcal H}\right)^2\Psi,
\end{equation}
where $\delta\rho_\mathrm{com}$ is the density perturbation in the comoving gauge and we used the Poisson equation. Therefore, $\zeta\simeq\mathcal R$ on super-horizon scales. Then, from Eqs.~\eqref{RR} and \eqref{zetazeta}, we obtain the following differential equation:
\begin{equation}
	\frac{\Psi'(\eta)}{\mathcal H(\eta)}+\frac{6+5r(\eta)}{2+2r(\eta)}\Psi(\eta)+\frac{4+3r(\eta)}{2+2r(\eta)}\zeta_\mathrm{inf}+\frac{r(\eta)}{2+2r(\eta)}S=0.\label{kitajima_de}
\end{equation}
The initial condition should be $\Psi(N\rightarrow-\infty)=-2\zeta_\mathrm{inf}/3$, where $N$ is the number of e-folds. 
Using the relation $r\propto e^N$ and defining a new function $\hat\Psi\equiv\Psi\exp\left(\int\mathrm dN\frac{6+5r}{2+2r}\right)=\frac{r^3}{\sqrt{1+r}}\Psi$, we can solve Eq.~\eqref{kitajima_de} analytically as
\begin{equation}
	\Psi=\frac{-9r^3-2r^2+8r+16-16\sqrt{1+r}}{15r^3}\zeta_\mathrm{inf}+\frac{-r^3+2r^2-8r-16+16\sqrt{1+r}}{5r^3}S.
\end{equation}
When $\zeta_\mathrm{inf}$ is negligible, it is rewritten as
\begin{eqnarray}
	\Psi(\eta)&=&-\frac{r(\eta)}{6+5r(\eta)}(1-\Delta_\sigma(\eta))S\qquad\mathrm{(superhorizon)}\label{superh}\\
	\Delta_\sigma(r(\eta))&\equiv&1+\frac{(6+5r)(-r^3+2r^2-8r-16+16\sqrt{1+r})}{5r^4}.
\end{eqnarray}

Next, we discuss the behavior of $\Psi$ on sub-horizon ($k\eta\gg1$) scales. $(00)$ component of the Einstein equation at linear order of perturbation is written as
\begin{equation}
	3\mathcal H(\Psi'+\mathcal H\Psi)+k^2\Psi=-\frac{1}{2M_P^2}a^2\delta\rho.
\end{equation}
On sub-horizon scales, the first term of the left-hand side is negligible and the Poisson equation is valid. In the axion-like curvaton model, it reads
\begin{equation}
	k^2\Psi=-\frac{1}{2M_P^2}a^2\delta\rho=-\frac{3}{2}\mathcal H^2\frac{1}{1+r}\frac{\delta\rho_r+\delta\rho_\sigma}{\rho_r}.
\end{equation}
It is rewritten as
\begin{equation}
	\Psi(\eta)=-\frac{3}{2}\left(\frac{\mathcal H(\eta)}{k}\right)^2\left(\frac{1}{1+r(\eta)}\frac{\delta\rho_r}{\rho_r}+\frac{r(\eta)}{1+r(\eta)}\frac{\delta\rho_\sigma}{\rho_\sigma}\right)\simeq-\frac{3}{2}\left(\frac{\mathcal H(\eta)}{k}\right)^2\frac{r(\eta)}{1+r(\eta)}S\qquad\mathrm{(subhorizon)}.\label{subh}
\end{equation}
Note that $\mathcal H=1/\eta$ in the radiation-dominated universe. Interpolating Eqs.~\eqref{superh} and \eqref{subh}, we obtain
\begin{equation}
	\Psi(\eta)=-\frac{r(\eta)}{6+5r(\eta)}(1-\Delta_\sigma(\eta))\left[1+\frac{2(1+r(\eta))}{3(6+5r(\eta))}(1-\Delta_\sigma(\eta))(k\eta)^2\right]^{-1}S.
\end{equation}
Then, the transfer function $T_S$ is derived as
\begin{equation}
	T_S(\eta,k)=-\frac{r(\eta)}{6+5r(\eta)}(1-\Delta_\sigma(\eta))\left[1+\frac{2(1+r(\eta))}{3(6+5r(\eta))}(1-\Delta_\sigma(\eta))(k\eta)^2\right]^{-1}.\label{T_S}
\end{equation}

In summary, the transfer function is given by Eqs.~\eqref{T_Psi} and \eqref{T_S}. The density parameter of GWs can be calculated from Eqs.~\eqref{P_h_Z} and \eqref{Omega_GW_Z}.

\subsection{Suppression via Non-Gaussianity}\label{subsec:NG_suppression}

In evaluating GWs, it is important to note that non-Gaussianity characterized by a positive $f_\mathrm{NL}$ suppresses GWs. The qualitative explanation is given in the following. PBHs come from the tail of the PDF of the perturbations while the density of GWs is determined by the variance. 
If we take a positive $f_\mathrm{NL}$ and make $\left<\zeta_g^2\right>$ fixed, the shape of the PDF is distorted and the tail with large fluctuation is lifted. Then, in order to fix the abundance of PBHs, $\beta$, the variance of the PDF should be smaller, which leads to smaller GWs.

Now, let us make the discussion above more quantitative~\cite{Young:2013oia,Nakama:2016gzw}. Non-Gaussianity is not considered in the previous discussions except for Sec.~\ref{subsec:non-Gaussianity} and we will derive the correction here.\footnote{
It may be more accurate to calculate $\beta$ and $\mathcal P_\zeta$ numerically by using a non-Gaussian PDF from the beginning of the discussion. However, in this paper, we estimate them more roughly by following a conventional procedure, where we search for the variance of the Gaussian part which reproduces a given $\beta$.
}
The PDF of the Gaussian part of Eq.~\eqref{quadratic_NG1}, $\zeta_g$, is written as
\begin{equation}
	P_G(\zeta)=\frac{1}{\sqrt{2\pi\left<\zeta_g^2\right>}}\exp\left(-\frac{\zeta^2}{2\left<\zeta_g^2\right>}\right).
\end{equation}
From Eq.~\eqref{quadratic_NG1}, there are two $\zeta_g$ corresponding to given $\zeta$:
\begin{equation}
	\zeta_{g\pm}(\zeta)=\frac{5}{6f_\mathrm{NL}}\left[-1\pm\sqrt{1+\frac{12f_\mathrm{NL}}{5}\left(\frac{3f_\mathrm{NL}\left<\zeta_g^2\right>}{5}+\zeta\right)}\right].
\end{equation}
The PDF of $\zeta$ including non-Gaussianity  is given by
\begin{eqnarray}
	P_{NG}(\zeta)&=&\sum_{i=\pm}\left|\frac{\mathrm d\zeta_{g,i}(\zeta)}{\mathrm d\zeta}\right|P_G(\zeta_{g,i}(\zeta)).
\end{eqnarray}
The production rate of PBHs at the formation, $\beta$, is expressed as\footnote{
As mentioned in Footnote~\ref{delta_not_zeta}, $\beta$ should be evaluated in terms of $\delta$, not $\zeta$. 
However, evaluation in terms of $\delta$ is non-trivial when non-Gaussianity is involved. 
Then, we follow the well-established method using $\zeta$. 
In that sense, the argument given here is a rough estimation. 
The uncertainty will be attributed to the evaluation of $\zeta_\mathrm{c}$ and the robust evaluation is left for a future work.
}
\begin{eqnarray}
	\beta&=&\int_{\zeta_\mathrm{c}}\mathrm d\zeta \,P_{NG}(\zeta)\\
	&=&\frac{1}{\sqrt{2\pi}}\left(\int_{y_{\mathrm{c}+}}^\infty\mathrm dy\,e^{-\frac{y^2}{2}}+\int_{-\infty}^{y_{\mathrm{c}-}}\mathrm dy\,e^{-\frac{y^2}{2}}\right),\label{beta_NG}
\end{eqnarray}
where
\begin{equation}
	y_{\mathrm{c}\pm}\equiv\frac{\zeta_{g\pm}(\zeta_\mathrm{c})}{\sqrt{\left<\zeta_g^2\right>}}
\end{equation}
and we can evaluate the threshold as  $\zeta_\mathrm{c}=\frac{9}{4}\delta_\mathrm{c}=0.9$ using the relation $\left<\delta_{\mathrm{cg}}^2\right>\simeq\frac{16}{81}\,\mathcal P_{\zeta}$ (see also the discussion given in Sec.~\ref{PBH_formation}).

On the other hand, $\beta$ is calculated by using Eqs.~\eqref{beta} and \eqref{16/81}, where non-Gaussianity is not involved. 
The height of the mass spectrum depends on $\beta$ through Eq.~\eqref{Omega_PBH} and so the value of $\beta$ should be fixed.
As we will see in Sec.~\ref{sec_Result}, the mass spectrum of $\Omega_\text{PBH}/\Omega_c$ has a low-mass cutoff, $M_\text{min}$ (see Fig.~\ref{fig:mass_spectrum}).
Since PBHs with around $M_\mathrm{min}$ dominantly contribute to $\Omega_\text{PBH}/\Omega_c$, we fix the value of $\beta(M_\text{min})$ and calculate the variance of the Gaussian part $\left<\zeta_g^2\right>$ which reproduces the value of $\beta(M_\text{min})$.

Now we are in a position to evaluate $\mathcal P_{\zeta,\mathrm{curv}}$ by using $\left<\zeta_g^2\right>$ which reproduces $\beta(M_\text{min})$.
As in the case of Eq.~\eqref{P_zeta_g_log}, 
where the power spectrum is scale-invariant,
 we can take $\left<\zeta_g^2\right>\simeq \mathcal P_{\zeta_g,\mathrm{curv}}(k_*)$ up to a logarithmic correction in the order of unity.
 In the following, we use this relation ($\left<\zeta_g^2\right>\simeq \mathcal P_{\zeta_g,\mathrm{curv}}(k_*)$) to get $P_{\zeta_g,\mathrm{curv}}(k_*)$ which reproduces $\beta(M_\text{min})$.
As shown in Eq.~\eqref{PNL_clean}, the variance of full $\zeta$ is given by  
\begin{equation}
	\mathcal P_{\zeta,\mathrm{curv}}(k)=\mathcal P_{\zeta_g,\mathrm{curv}}(k)+\left(\frac{3}{5}f_\mathrm{NL}\right)^2\frac{k^3}{2\pi}\int\mathrm d^3p\frac{1}{p^3}\frac{1}{\left|\bvec k-\bvec p\right|^3}\mathcal P_{\zeta_g,\mathrm{curv}}(p)\mathcal P_{\zeta_g,\mathrm{curv}}(|\bvec k-\bvec p|).
\end{equation}
We want to evaluate the integral in the case of the axion-like curvaton model by using Eqs.\eqref{Pzeta_flat} and \eqref{Pzeta_blue}. It is well approximated for $k$ larger than $k_*$ as\footnote{
The PTA constraints exist for $k>k_*$ (see Sec.\ref{sec_Result}). 
}
\begin{equation}
	\mathcal P_{\zeta,\mathrm{curv}}(k)\simeq\mathcal P_{\zeta_g,\mathrm{curv}}(k_*)+4\left(\frac{3}{5}f_\mathrm{NL}\right)^2
	\left[\frac{1}{n_\sigma-1}+\ln\left(\frac{k}{k_*}\right)\right]\mathcal P_{\zeta_g,\mathrm{curv}}^2(k_*)
	\quad \mathrm{for}\quad k>k_*
	.
\end{equation}
This expression is similar to Eq.~\eqref{zeta_k}, where $\mathcal P_{\zeta_g,\mathrm{curv}}$ is scale-invariant, with $L^{-1}$ replaced by $k_*$. Neglecting the logarithmic correction, we obtain
\begin{equation}
	\mathcal P_{\zeta,\mathrm{curv}}(k)\simeq\mathcal P_{\zeta_g,\mathrm{curv}}(k_*)+4\left(\frac{3}{5}f_\mathrm{NL}\right)^2
	\mathcal P_{\zeta_g,\mathrm{curv}}^2(k_*)
	\quad \mathrm{for}\quad k>k_*
	.\label{Pzeta_order_estimate}
\end{equation}
Plugging $f_\mathrm{NL}$ and $\mathcal P_{\zeta_g,\mathrm{curv}}$ into Eq.~\eqref{Pzeta_order_estimate}, we obtain the power spectrum of full $\zeta$, $\mathcal P_{\zeta,\mathrm{curv}}$.

On the other hand, the formula for $\mathcal P_{\zeta,\mathrm{curv}}(k_*)$ given in Eq.~\eqref{Pzeta_flat} does not involve non-Gaussianity. 
Then, if one calculates $\mathcal P_h$ given in Eq.~\eqref{P_h} by using $\mathcal P_{\zeta,\mathrm{curv}}(k_*)$ given in Eq.~\eqref{Pzeta_flat}, a suppression factor should be multiplied for more realistic evaluation of $\mathcal P_h$.
We denote the suppression factor as $Q$. Taking care of $\mathcal P_h\sim\mathcal P_\zeta^2$, the factor $Q$ is given by
\begin{equation}
	Q=\left(\frac{\mathcal P_{\zeta,\mathrm{curv}}(k_*)}{\left.\mathcal P_{\zeta,\mathrm{curv}}(k_*)\right|_{f_\mathrm{NL}=0}}\right)^2\label{Q}.
\end{equation}
Here, the numerator, $\mathcal P_{\zeta,\mathrm{curv}}(k_*)$, stands for the power spectrum of full $\zeta$ and is given by Eq.~\eqref{Pzeta_order_estimate} while the denominator, $\left.\mathcal P_{\zeta,\mathrm{curv}}(k_*)\right|_{f_\mathrm{NL}=0}$, stands for the variance which would give the same $\beta$ as $\mathcal{P}_{\zeta,\text{curv}}(k_*)$ if non-Gaussianity was not taken into account. 

\section{Result}\label{sec_Result}

In this section, we show that there is a set of parameters for the axion-like curvaton model which reproduces the merger event rate estimated by the LIGO-Virgo Collaboration, $12$--$213$Gpc$^{-3}$yr$^{-1}$~\cite{Abbott:2017vtc}.

\subsection{Mass Spectrum for the LIGO events}

\begin{figure}
	\centering
	\includegraphics[width=.45\textwidth]{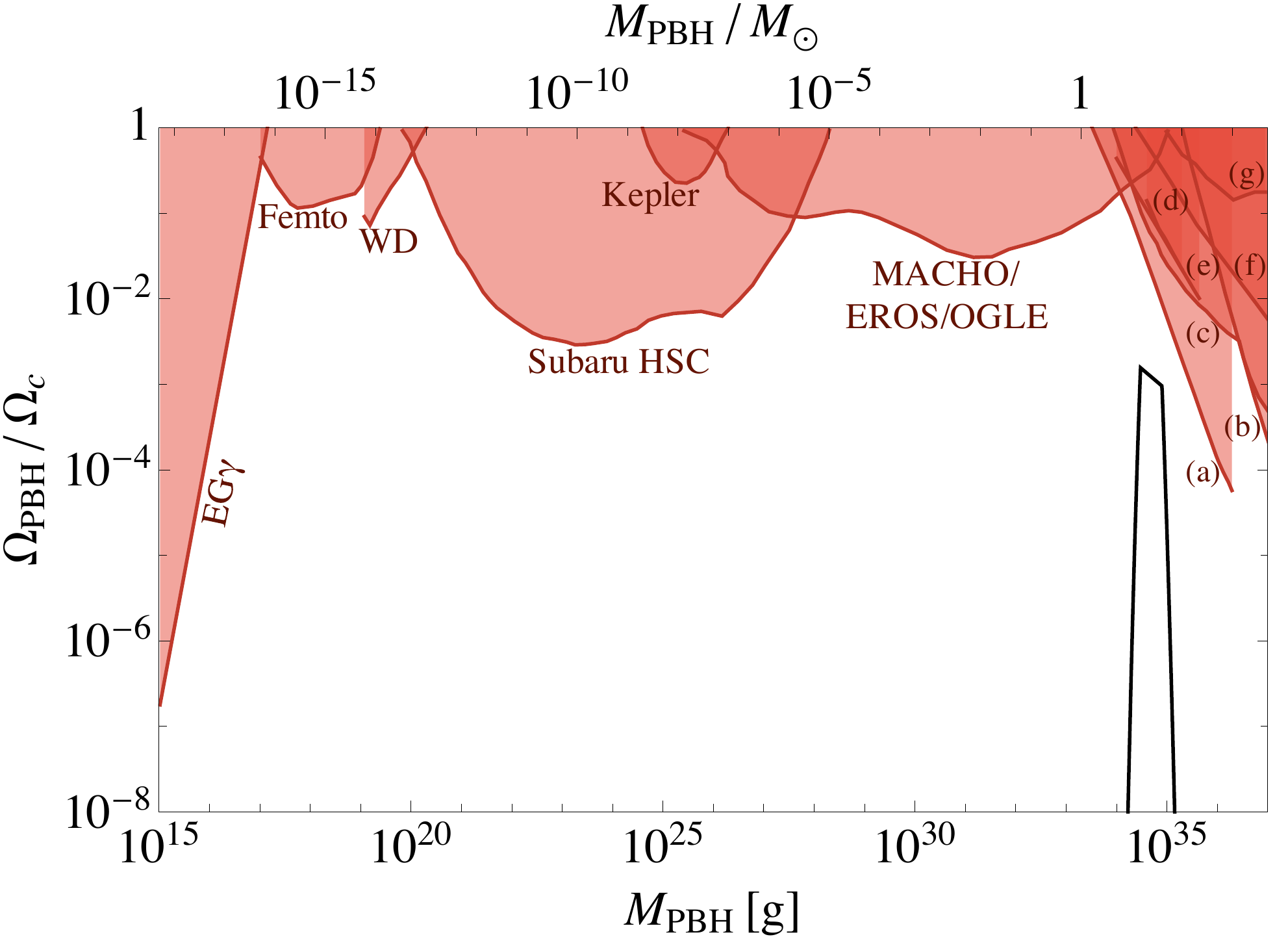}
	\caption{\small
	{\bf Black solid line:} the PBH mass spectrum for parameters given in Eq.~\eqref{parameter}. 
	{\bf Red shaded regions:} observationally excluded regions for a monochromatic mass function.\footnote{
		The constraints for extended mass functions are discussed in
		\cite{Carr:2016drx,
		Green:2016xgy,
		Inomata:2017okj,
		Kuhnel:2017pwq,
		Carr:2017jsz,
		Bellomo:2017zsr,
		Inomata:2017bwi}.
		}
	Observational targets are as follows.
	\emph{EG$\gamma$}: the extragalactic gamma-ray produced through Hawking radiation of PBHs~\cite{Carr:2009jm}.
	\emph{Femto}: the femtolensing of gamma-ray burst~\cite{Barnacka:2012bm}.
	\emph{WD}: the survival of the white dwarf named RX J0648.0-4418~\cite{Graham:2015apa}.
	\emph{Subaru HSC, Kepler, EROS/MACHO/OGLE}: the microlensing of stars due to PBHs with Subaru HSC~\cite{Niikura:2017zjd}, Kepler satellite~\cite{Griest:2013esa} and EROS/MACHO/OGLE ~\cite{Allsman:2000kg,Tisserand:2006zx,Wyrzykowski:2011tr}.
	\emph{(a,b)}: the variations of CMB spectrum caused by the radiation which originates from the gas accretion onto PBHs~\cite{Poulin:2017bwe,Ali-Haimoud:2016mbv}.
	\emph{(c,d)}: the radio and X-ray from accretion~\cite{Inoue:2017csr,Gaggero:2016dpq}.
	\emph{(e,f)}: the dynamical heating of dwarf galaxies~\cite{Koushiappas:2017chw} and ultra-faint dwarf galaxies~\cite{Brandt:2016aco}.
	\emph{(g)}: the distribution of wide binaries~\cite{Monroy-Rodriguez:2014ula}.
	The constraint from Subaru HSC would be weaker than that shown here for $M_\mathrm{PBH}\lesssim10^{-10}M_\odot$ due to the wave effect~\cite{Niikura:2017zjd
	}. 
	Besides the constraints shown here, the BH-BH merger rate currently estimated by LIGO indicates that $\Omega_\mathrm{PBH}/\Omega_c\lesssim10^{-1.5}$ for $M_\mathrm{PBH}\sim30M_\odot$~\cite{Raidal:2017mfl}.
	}
	\label{fig:mass_spectrum}
\end{figure}

We take the model parameters as
\begin{eqnarray}
	&&\kappa=1.0\times10^{-2},\quad
	m_\sigma=5.5\times10^3\,\mathrm{GeV},\quad
	f=1.9\times10^{13}\,\mathrm{GeV},\quad
	n_\sigma=2.5,\quad\nonumber\\&&
	\theta=0.3,\quad
	\varphi(k_p)=2.44\times10^{18}\,\mathrm{GeV},\quad
	H_\mathrm{inf}=3.7\times10^{13}\,\mathrm{GeV},\quad
	\Gamma_I=3.8\times10^2\,\mathrm{GeV},
	\label{parameter}
\end{eqnarray}
where $k_p=0.05\,\mathrm{Mpc^{-1}}$ is the pivot scale.
The mass spectrum of PBHs is calculated through Eq.~\eqref{Omega_PBH} and shown in Fig.~\ref{fig:mass_spectrum}. The height is normalized by the abundance of cold dark matter. The spectrum has a peak at $\sim\,30M_\odot$ with the height of $10^{-3}$.
It was shown that the merger event rate estimated by the LIGO-Virgo Collaboration is reproduced if $\Omega_\mathrm{PBH}/\Omega_\mathrm{c}$ is in the order of $10^{-3}$~\cite{Sasaki:2016jop}.
Then, the set of parameters given in Eq.~\eqref{parameter} reproduces the LIGO events.

We will give some explanation for the shape of the spectrum in Fig.~\ref{fig:mass_spectrum}.
The steep damping at the right side of the peak is due to the blue-tilted region of the generated curvature perturbation. The cutoff of the left side of the peak is the minimum mass of PBH $M_\mathrm{min}$. Since PBHs are formed dominantly after the curvaton decays, $M_\mathrm{min}=M(k_\mathrm{dec})$ (see Eq.\eqref{PBHmass}).

For further understandings of the scenario, we will give some remarks. With the help of the argument given in Sec.~\ref{axion_curvaton_model} and Sec.~\ref{PBH_formation}, we can relate the parameters given in Eq.~\eqref{parameter} to other quantities such as
\begin{eqnarray}
	&&k_\mathrm{c}=6.8\,\mathrm{Mpc^{-1}},\quad
	k_*=3.2\times10^{5}\,\mathrm{Mpc^{-1}},\quad
	k_\mathrm{dec}=5.2\times10^{5}\,\mathrm{Mpc^{-1}},\quad
	M_*=8.0\times10^{34}\,\mathrm{g},\quad
	M_\mathrm{min}=3.0\times10^{34}\,\mathrm{g},\quad
	\nonumber\\&&
	T_\mathrm{dec}=45\,\mathrm{MeV},\quad
	T_R=1.6\times10^{10}\,\mathrm{GeV},\quad
	r_D=0.35,\quad
	f_\mathrm{NL}=4.17,\quad
	g_*(t_\mathrm{dec})=10.75,\quad
	g_*(t_R)=106.75,
	\nonumber\\&&
	\left.\mathcal P_{\zeta,\mathrm{curv}}(k_*)\right|_{f_\mathrm{NL}=0}=2.01\times10^{-2},\quad
	\left<\delta_{\mathrm{cg}}^2(M_\mathrm{min})\right>=3.97\times10^{-3},\quad
	\beta(M_\mathrm{min})=1.1\times10^{-10}
	\label{parameter2}.
\end{eqnarray}

There are cosmological conditions to be satisfied for the consistency of the model.
 First, as already mentioned in Eq.~\eqref{1Mpc}, $k_\mathrm{c}$ should be larger than $1\,\mathrm{Mpc^{-1}}$ so that the inflaton dominates over the curvaton on scales larger than $1\,\mathrm{Mpc}$.
  Next, $T_\mathrm{dec}$ should be higher than $\mathcal O(1)\,\mathrm{MeV}$ in order for the abundance of $^4 \text{He}$ to be consistent with the observations~\cite{Kawasaki:1999na,Kawasaki:2000en}. 
  These conditions are satisfied by the parameters given in Eq.~\eqref{parameter2}. 
 From Eq.~\eqref{parameter2}, we can see that $\mathcal P_\zeta\sim10^{-2}$ is needed in order for a relevant abundance of PBHs to be produced. 
  For such large fluctuations, $H_\mathrm{inf}$ and $f$ are expected to be comparable because of Eq.~\eqref{Pzeta_flat}, which is the case in Eq.~\eqref{parameter}. 
  We also make a remark about parameters different from Eq.~\eqref{parameter}. 
  As far as a spectrum similar to Fig.~\ref{fig:mass_spectrum} is generated, smaller $H_\mathrm{inf}$ corresponds to larger $T_R$ because of Eqs.~\eqref{rD1} and \eqref{Pzeta_flat}.
  Then, too small $H_\mathrm{inf}$ is impossible because the energy scale during the inflation should be larger than that at the reheating, which leads to  $H_\mathrm{inf}\gtrsim10^{11}\,\mathrm{GeV}$.
  Also, so large $H_\mathrm{inf}$ is not preferable because it leads to large $\varphi(k_p)$, which is expected not to exceed the Planck scale so much.
  At the same time, there is a constraint from the Planck observation~\cite{Ade:2015xua}: $H_\mathrm{inf}<5\times10^{-5}M_P$.

The parameter $\theta$ given in Eq.~\eqref{parameter} is taken carefully. Although we approximate the curvaton potential as quadratic, it is actually of the cosine type. Then, $\theta+\delta\theta<\pi$ must be satisfied at least when $\delta\theta$ is the fluctuation of the misalignment angle corresponding to the threshold $\delta_\mathrm{c}$. Since $\left<\delta_{\mathrm{cg}}^2\right>\simeq \frac{16}{81}\,\mathcal P_\zeta\simeq\frac{16}{81}\left(\frac{r_D}{4+3r_D}\frac{2\delta\theta}{\theta}\right)^2$, the condition becomes $\theta+\delta\theta\simeq\left(1+\frac{9}{4}\frac{4+3r_D}{2r_D}\delta_\mathrm{c}\right)\theta<\pi$. In the case of $r_D=0.35$, it reads $\theta<0.42$.

\subsection{GWs and PTA Experiments}

We need to consider primordial GWs for the consistency of the model.
The frequency of GWs and the mass of  PBHs are related through Eq.~\eqref{PBHmass}.
 In the frequency region associated with $\sim M_\odot$, there are constraints on GWs from the observation of the pulsar timing array (PTA)~\cite{Lentati:2015qwp,Shannon:2015ect,Arzoumanian:2015liz,Moore:2014lga,Janssen:2014dka}. 
 Thus, we calculated the GW spectrum induced by the second-order effect and compare it with the PTA constraints.

Following the procedure shown in Sec.~\ref{subsec:NG_suppression}, we find that $\beta(M_\mathrm{min})=1.1\times10^{-10}$ given in Eq.~\eqref{parameter2} is reproduced when the variance of the Gaussian part of $\zeta$ takes the value of $\left<\zeta_g^2\right>=\mathcal{P}_{\zeta_g,\text{curv}}=4.7\times10^{-3}$. Then, from Eqs.~\eqref{Pzeta_order_estimate} and \eqref{Q}, the suppression factor is $Q=6.9\times10^{-2}$.

\begin{figure}
	\centering
	\includegraphics[width=.45\textwidth]{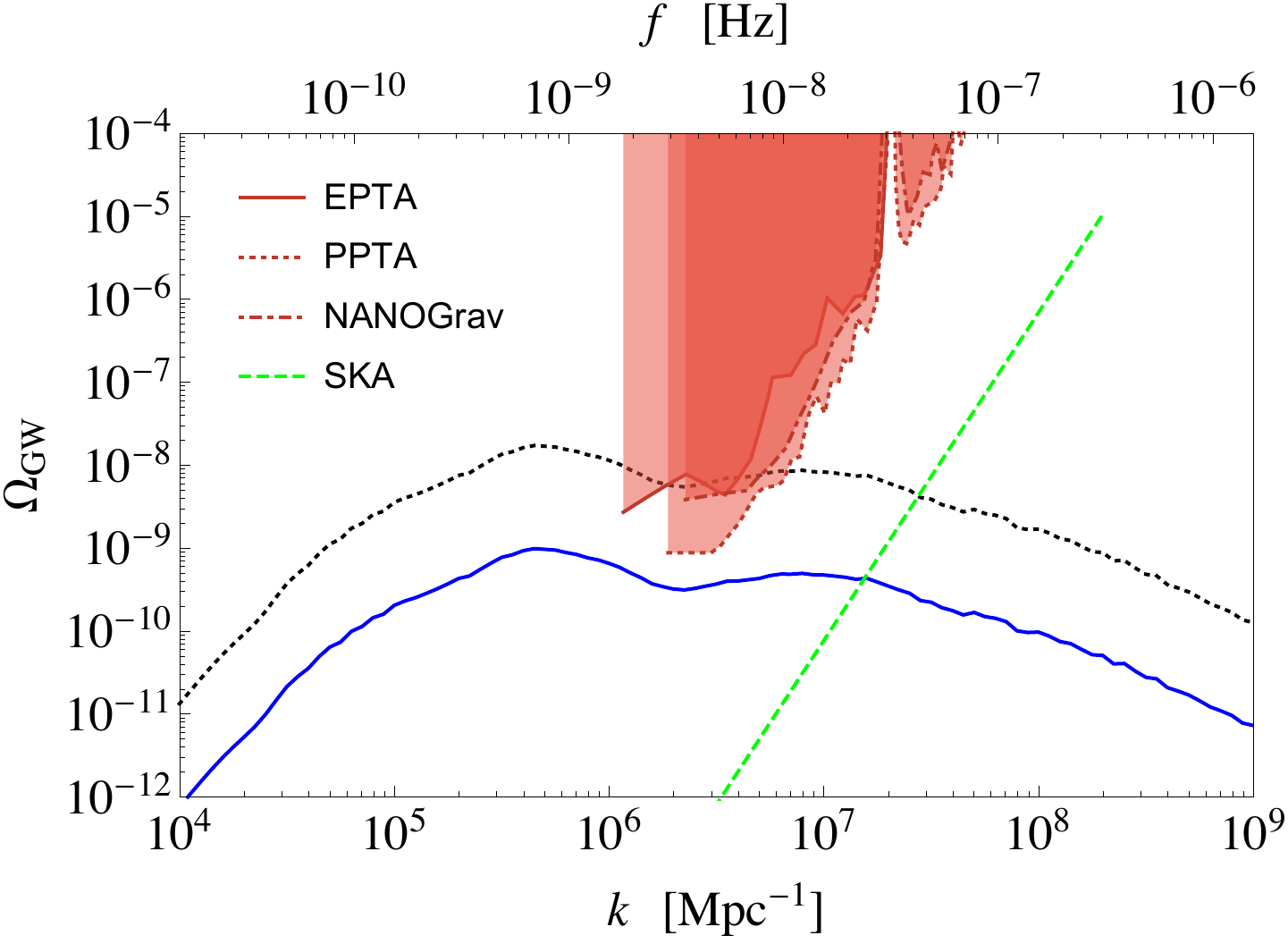}
	\caption{\small
	{\bf Blue solid line:} the GW spectrum for parameters given in Eq.~\eqref{parameter}. 
	{\bf Red shaded region:} the constraints from the PTA experiments with EPTA~\cite{Lentati:2015qwp}, PPTA~\cite{Shannon:2015ect}, and NANOGrav~\cite{Arzoumanian:2015liz}. 
	{\bf Green dashed line:} the constraints from the future SKA experiments~\cite{Moore:2014lga,Janssen:2014dka}.
	{\bf Black dotted line:} the GW spectrum for parameters given in Eq.~\eqref{parameter} with the amplitude changed so that the same $\beta$ is realized when we assume $f_\mathrm{NL}=0$.
	In other words, the amplitude of the black dotted line is larger than that of the blue solid line by the factor of $Q^{-1}$.
	}
	\label{fig:GW_spectrum}
\end{figure}

Using Eqs.~\eqref{P_h}, \eqref{Omega_GW} and the factor $Q$, we calculate the GW spectrum for parameters given in Eq.~\eqref{parameter} as shown in Fig.~\ref{fig:GW_spectrum}.
From this figure, we can see that the GW spectrum is marginally consistent with the constraints from the PTA experiments.
Due to the uncertainty discussed in Sec.~\ref{PBH_formation}, the spectrum could avoid the constraints more safely for a larger value of $\vev{\delta_\text{cg}^2(k^{-1})}/ \mathcal P_{\zeta,\text{curv}}(k,\eta=k^{-1})$.
For comparison, we also show the GW spectrum in the case where we do not take account of non-Gaussianity but the same $\beta$ is realized with a black dotted line.
We can also see that the non-Gaussianity plays an important role for the GW spectrum to avoid the constraints.

\begin{figure}[htbp]
  \begin{center}
    \begin{tabular}{c}

      \begin{minipage}{0.45\textwidth}
        \begin{center}
          \includegraphics[width=\hsize]{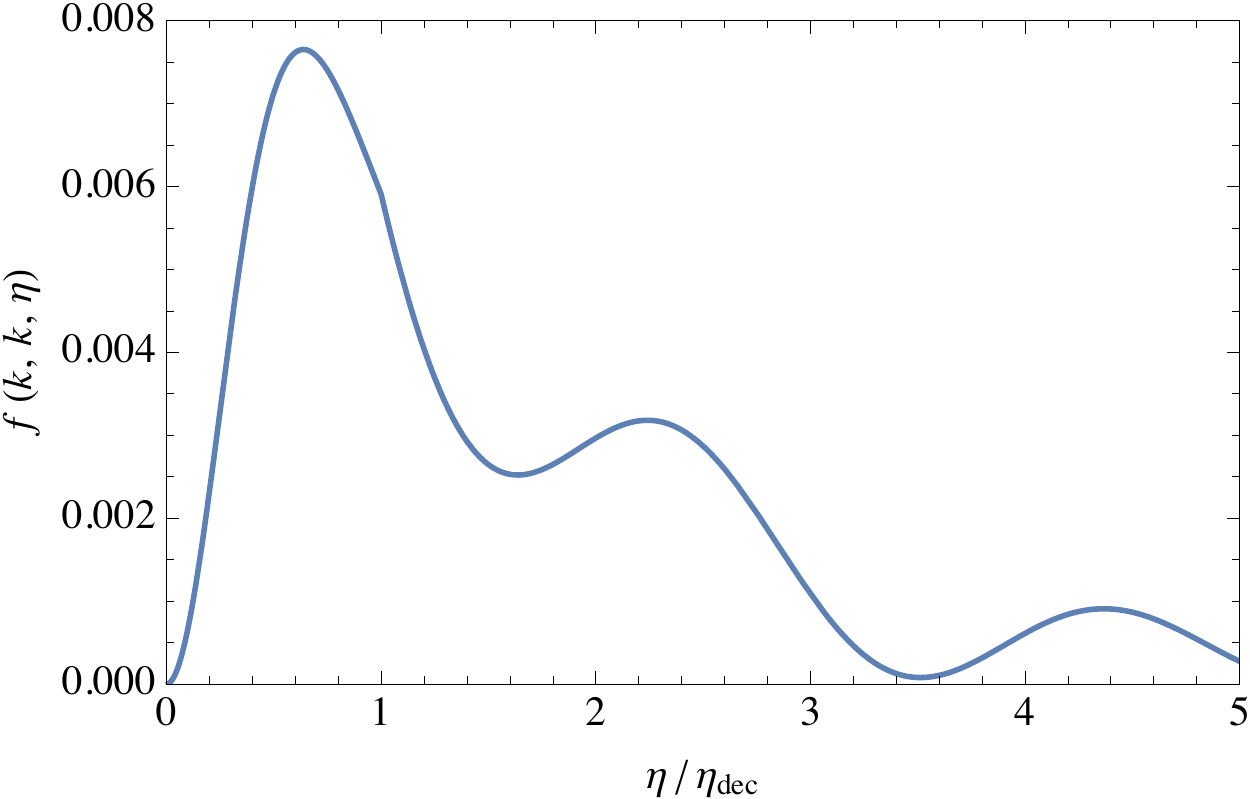}
          \hspace{1.6cm} 
          {\small (a) $k=3k_\mathrm{dec}=1.6\times10^6\,\mathrm{Mpc^{-1}}$}
        \end{center}
      \end{minipage}
	\hspace{0.5cm} 
      \begin{minipage}{0.45\textwidth}
        \begin{center}
          \includegraphics[width=\hsize]{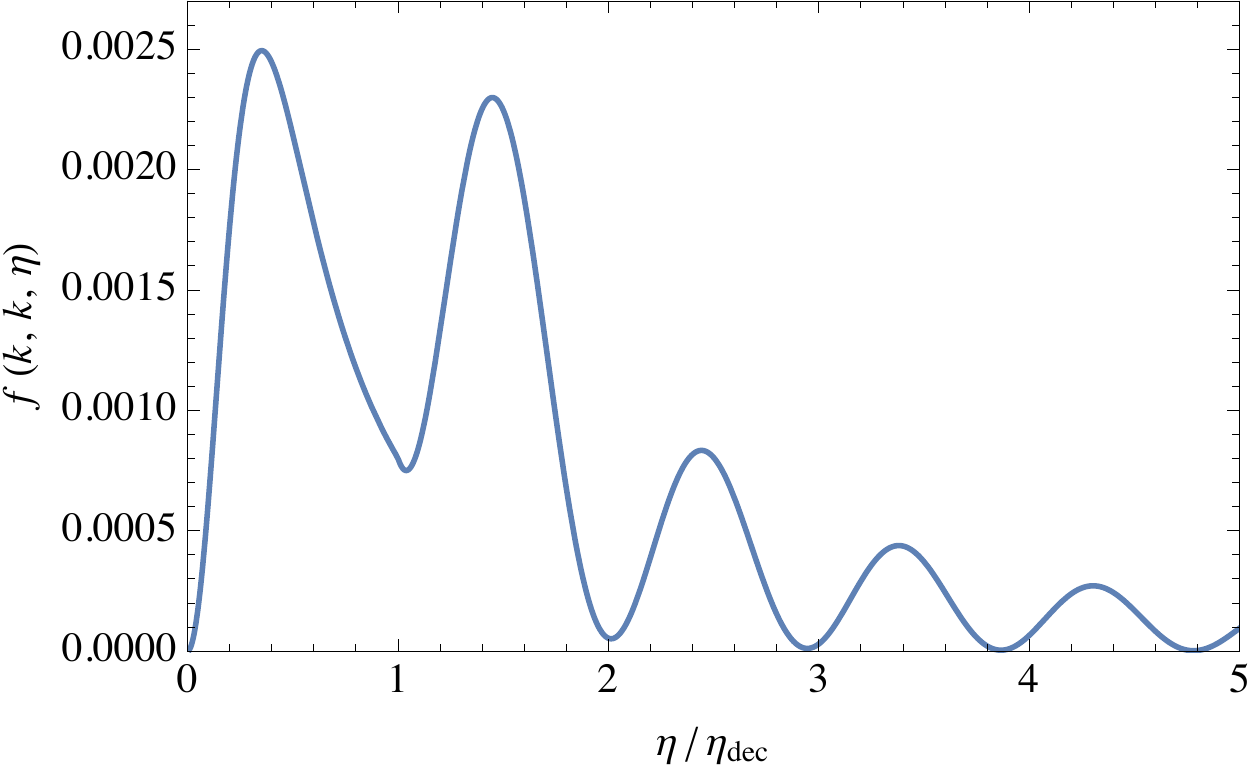}
          \hspace{1.6cm} 
          {\small (b) $k=6k_\mathrm{dec}=3.1\times10^6\,\mathrm{Mpc^{-1}}$}
        \end{center}
      \end{minipage}\\

      \begin{minipage}{0.45\textwidth}
      \vspace{0.6cm}
        \begin{center}
          \includegraphics[width=\hsize]{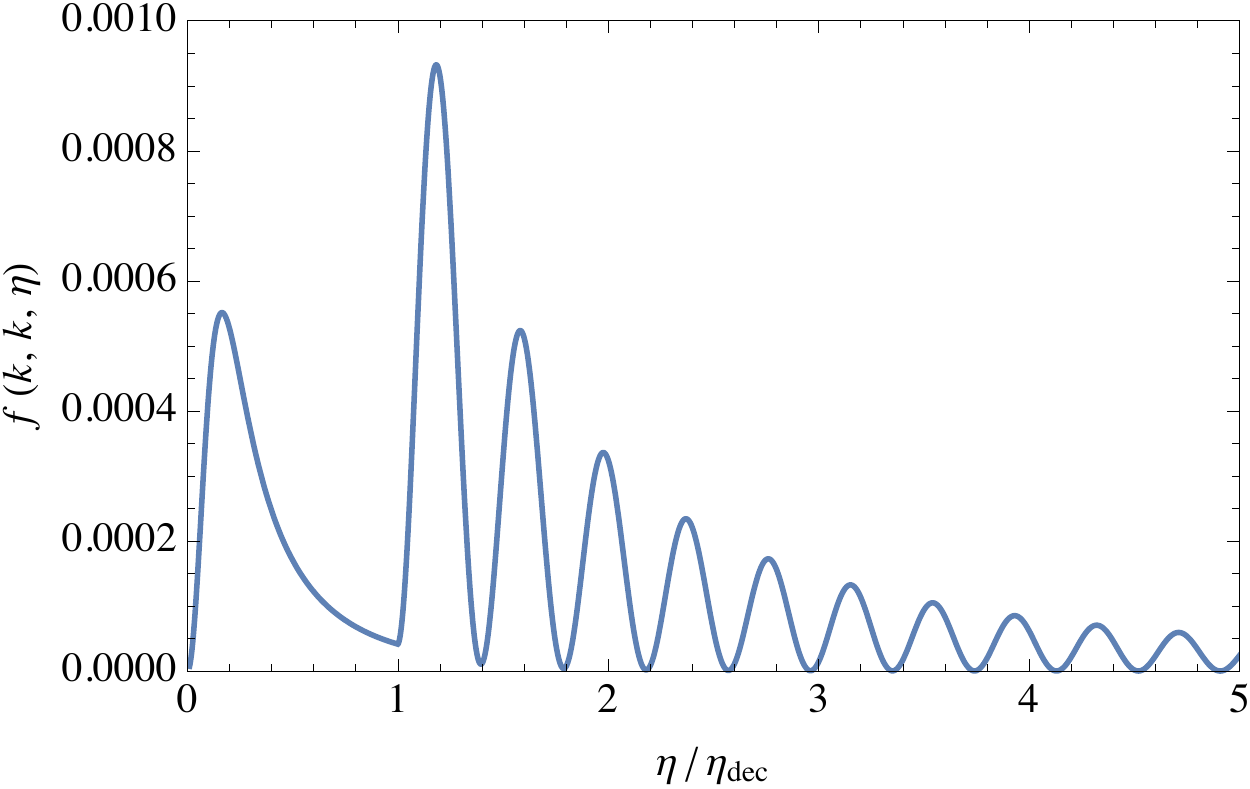}
          \hspace{1.6cm} 
          {\small (c) $k=14k_\mathrm{dec}=7.3\times10^6\,\mathrm{Mpc^{-1}}$}
        \end{center}
      \end{minipage}

    \end{tabular}
    \caption{\small	
    	$f(k,k,\eta)$ for each $k$ mode and parameters given in Eq.~\eqref{parameter}. 
	The function $f(k,k,\eta)$ stands for the time dependence of the source of GWs.
	In Fig.~(a), the peak at the horizon reentering is dominant.
	In Fig.~(c), the peak soon after the curvaton decays is dominant.
	In Fig.~(b), they are comparable.
	}
    \label{fig:f}
  \end{center}
\end{figure}

Now, we will give a physical interpretation of the shape of the GW spectrum. 
Basically, the GWs are dominantly produced at the horizon reentry of the perturbations~\cite{Saito:2008jc,Saito:2009jt}.
Then, the GW spectrum is likely to trace the curvature perturbation spectrum at the horizon reentry of each mode. 
This statement agrees with Fig.~\ref{fig:curvature_spectrum2} and Fig.~\ref{fig:GW_spectrum} for scales $k\lesssim k_\mathrm{dec}=5.2\times10^{5}\,\mathrm{Mpc^{-1}}$. 
However, a more explanation is needed for scales $k\gtrsim k_\mathrm{dec}$.
Actually, in the curvaton model, a significant amount of GWs are also produced soon after the curvaton decays.
Furthermore, for large $k$ modes, the GWs produced at the curvaton decay are larger than that produced at the horizon reentry. 
This can be checked by investigating the function $f(k_1,k_2,\eta)$, which stands for the time dependence of the source of GWs, given in Eq. \eqref{f} (see Fig.~\ref{fig:f}).
Thus, for $k\gtrsim k_\mathrm{dec}$, the GW spectrum shown in Fig.~\ref{fig:GW_spectrum} decreases more slowly as $k$ than expected from Fig.~\ref{fig:curvature_spectrum2}.
In particular, the spectrum has a small bump around $k\sim 10k_\text{dec}$.

\subsection{Curvaton decay}

Finally, let us discuss the preferred value of the coupling constant between the curvaton and the Standard-Model (SM) particles, $\kappa$. 
Assume that the interaction is induced by the anomaly of the gauge symmetries such as $SU(3)_C$ or $U(1)_{EM}$ similarly to the axion. 
The effective Lagrangian may contain the interaction as
\begin{equation}
	\Delta\mathcal L\sim\frac{g^2}{32\pi^2}\frac{\sigma}{f}F^{\mu\nu}\tilde F_{\mu\nu},
	\label{Lagrangian1}
\end{equation}
where $g$ is a coupling constant, $F_{\mu\nu}$ is the field strength of gluon or photon, $\tilde F_{\mu\nu}\equiv\frac{1}{2}\epsilon_{\mu\nu\rho\sigma}F^{\rho\sigma}$ is its dual and the exact coefficient depends on models.
This expression can be rewritten as
\begin{equation}
	\Delta\mathcal L\sim\frac{\alpha}{8\pi}\frac{\sigma}{f}F^{\mu\nu}\tilde F_{\mu\nu}=\frac{\alpha}{2\pi}\frac{\sigma}{f}\frac{1}{4}F^{\mu\nu}\tilde F_{\mu\nu}=-\frac{\alpha}{2\pi}\frac{\sigma}{f}\bvec E\cdot \bvec B,
	\label{Lagrangian2}
\end{equation}
where $\alpha\equiv g^2/4\pi$ is the fine structure constant and we assume the interaction with photon in the last equality.
Then, the decay rate of the curvaton is evaluated as
\begin{equation}
	\Gamma_\sigma\sim\frac{1}{64\pi}\left(\frac{\alpha}{2\pi f}\right)^2m_\sigma^3=\frac{\alpha^2}{256\pi^3}\frac{m_\sigma^3}{f^2}=\frac{1}{16\pi}\left(\frac{\alpha}{4\pi}\right)^2\frac{m_\sigma^3}{f^2}\label{decay_rate_alpha}
\end{equation}
Comparing Eqs.~\eqref{decay_rate} and \eqref{decay_rate_alpha}, one finds $\kappa\sim\alpha/4\pi$.
Then, it seems natural if $\kappa\sim10^{-3}\mathchar`-10^{-2}$.
The value $\kappa=1.0\times10^{-2}$ given in Eq.~\eqref{parameter} agrees with it. 
However, the decay of the curvaton may be caused by other interactions and the argument given above indeed depends on the physics beyond the SM.

\section{Conclusion}
\label{sec_Conclusion}

In this paper, we have revisited PBH formation in the axion-like curvaton model~\cite{Kawasaki:2012wr} 
and shown that this model can produce a significant amount of PBHs that account for the event rate of BHs merger estimated by the LIGO-Virgo Collaboration.
There are two main improvements.
The first one is related to the calculation of PBH production rate.
Following the analysis in Ref.~\cite{Young:2014ana}, we have calculated the production rate of PBHs in terms of density perturbations instead of curvature perturbations which was used in~\cite{Kawasaki:2012wr}.
The revised PBH mass spectrum can be broad between the mass scales corresponding to $k_*$ and $k_\text{dec}$. 
This contrasts with the spectrum  obtained in~\cite{Kawasaki:2012wr} which has a sharp peak at the mass scale corresponding to $k_\text{dec}$.  
The second one is related to the non-Gaussianity.
We have taken account of the non-Gaussianity produced in this model and discussed the effect of the non-Gaussianity on the GW spectrum quantitatively.
As a result, we have found that the non-Gaussianity plays an important role for the GW spectrum to avoid the PTA constraints in the case where the produced PBHs explain the LIGO events.
Also, we have seen that non-Gaussianity is helpful for the curvature spectrum to avoid the constraints from the $\mu$-distortion and BBN.

Our calculation given in this paper includes several uncertainties. 
First, there are uncertainties on PBH formation. 
We have taken $\delta_c=0.4$ and $\gamma=0.2$ but these parameters have uncertainties.
Also, the relation between the PBH abundance and the power spectrum of primordial perturbations depends on the choice of a window function.
Especially, if we adopt a real-space top-hat window, the PTA experiments impose no constraint on LIGO PBHs~\cite{Ando:2018qdb}.
We have taken $\vev{\delta_\mathrm{cg}^2}=0.2\mathcal P_\zeta$ as the fiducial relation. 

Next, our analysis on how non-Gaussianity changes the power spectrum required for a fixed amount of PBHs is a rough estimation.
We stop our calculation with $f_\mathrm{NL}$ i.e. quadratic non-Gaussianity.
For example, in Ref.~\cite{Young:2013oia}, a method to calculate the non-linear effect of non-Gaussinity is demonstrated by using curvature perturbations for the quadratic curvaton.
However, following this method will not necessarily make our analysis more accurate due to the following uncertainties:
1) The PBH abundance should be evaluated by using coarse-grained density perturbations as mentioned above but the non-linear relation between density/curvature perturbations is non-trivial.
2) The quadratic curvaton is only an approximation in our case as mentioned below.
3) The effect of non-Gaussianity is degenerate with the uncertainties on the choice of a window function.

Last, throughout this paper, the cosine-type potential of the curvaton is approximated by a quadratic potential as in Eq.~\eqref{curvaton_potential}.
The deviation from this approximation is left for a future work. 
It may also be a good target of an application of the stochastic-$\delta N$ formalism~\cite{Enqvist:2008kt,Fujita:2013cna,Fujita:2014tja,Vennin:2015hra,Kawasaki:2015ppx,Assadullahi:2016gkk,Vennin:2016wnk}.
There are some studies of axionic curvatons~\cite{Kawasaki:2011pd,Kobayashi:2012ba,Kawasaki:2012gg,Hardwick:2017fjo}
that suggest that the deviation from the quadratic potential may lead to larger $\mathcal P_\zeta$ and smaller $f_\mathrm{NL}$.
It is guessed that the variation of $\mathcal P_\zeta$ and $f_\mathrm{NL}$ may work as if they are compensating for each other since smaller $f_\mathrm{NL}$ requires larger $\mathcal P_\zeta$ to produce the same amount of PBHs.

The axion-like curvaton  model is built in the framework of SUSY.
Here we make a comment on the relation between SUSY breaking scale and the curvaton potential. 
For successful formation of the LIGO PBHs, the mass and the breaking scale of the axion-like curvaton are $m_\sigma  \sim  10^4$~GeV and $f\sim 10^{13}$~GeV, respectively. 
This suggests the curvaton scale $\Lambda \sim \sqrt{m_\sigma f} \sim 10^{8\mathchar`-9}\mathrm{GeV}$. If we identify the scale $\Lambda$ with the dynamical scale for SUSY breaking, the gravitino mass is given by $m_{3/2}\sim\Lambda^2/(4\pi M_\mathrm{Pl}) \sim 1\,\mathrm{MeV}\mathchar`-1\,\mathrm{GeV}$. It is known that gauge mediation models with such small gravitino masses are consistent with all observations~\cite{Hamaguchi:2014sea}. Moreover, the gravitino of mass $O(1)\, \mathrm{MeV}\mathchar`-O(1) \,\mathrm{GeV}$ is a good candidate for the dark matter~\cite{Buchmuller:2006nj}.

\section*{Acknowledgements}
\small\noindent
This work is supported by Grants-in-Aid for Scientific Research from the Ministry of Education,
Science, Sports, and Culture (MEXT), Japan,  No.\ 15H05889 (M.K.), No.\ 17H01131 (M.K.), No.\ 17K05434 (M.K.),
No.\ 26104009 (T.T.Y.), No.\ 26287039 (T.T.Y.), and No.\ 16H02176 (T.T.Y.), 
a World Premier International Research Center Initiative (WPI Initiative), MEXT, Japan 
(K.A., K.I., M.K., K.M., and T.T.Y.),
JSPS Research Fellowship for Young Scientists Grants No. 18J21906 (K. A.), No. 18J12728 (K. I.), and No. 15J07034 (K. M.),
and Advanced Leading Graduate Course for Photon Science (K.A., K.I.).

\normalsize

\appendix
\section{GWs induced by scalar perturbations}
\label{app_GW}

In this appendix, we review the formalism for evaluating GWs induced by scalar perturbations~\cite{Ananda:2006af,Baumann:2007zm,Saito:2008jc,Saito:2009jt,Bugaev:2009zh,Bugaev:2010bb,Inomata:2016rbd}.
 Since we intend to apply it to the curvaton model in Sec.~\ref{sec_GW}, we assume the case where isocurvature perturbations are translated into curvature perturbations.
We take the $(-+++)$ convention for the metric of the spacetime. The metric perturbation in the conformal Newtonian gauge is written as
\begin{equation}
	\mathrm ds^2=-a^2(1+2\Phi)\mathrm d\eta^2+a^2\left[(1-2\Psi)\delta_{ij}+\frac{1}{2}h_{ij}\right]\mathrm dx^i\mathrm dx^j.
\end{equation}
Here, we use $\Psi$ to indicate the curvature perturbation in the conformal Newtonian gauge.
The GWs $h_{ij}$ are transverse and traceless: $\partial_i h_{ij}=h_{ii}=0$. We assume that the anisotropic stress-energy tensor is negligible, which leads to $\Phi=\Psi$~\cite{dodelson,mukhanov}.

\subsection{Equation of motion of GWs}

It is useful to expand $h_{ij}$ in terms of the two polarization bases as
\begin{equation}
	h_{ij}(\eta,\bvec x)=\int\frac{\mathrm d^3k}{(2\pi)^3}\left[h_{\bvec k}^{(+)}(\eta)e_{ij}^{(+)}(\bvec k)+h_{\bvec k}^{(\times)}(\eta)e_{ij}^{(\times)}(\bvec k)\right]e^{i\bvec k\cdot \bvec x}.
\end{equation}
The two polarization bases are defined as
\begin{eqnarray}
	e_{ij}^{(+)}(\bvec k)&=&\frac{1}{\sqrt 2}\left[e_i(\bvec k)e_j(\bvec k)-\bar e_i(\bvec k)\bar e_j(\bvec k)\right]\\
	e_{ij}^{(\times)}(\bvec k)&=&\frac{1}{\sqrt 2}\left[e_i(\bvec k)\bar e_j(\bvec k)+\bar e_i(\bvec k) e_j(\bvec k)\right],
\end{eqnarray}
where $e_i(\bvec k)$ and $\bar e_i(\bvec k)$ are two orthonormal vectors orthogonal to $\bvec k$ and the polarization bases are normalized as $e_{ij}^{(+)}e_{ij}^{(+)}=e_{ij}^{(\times)}e_{ij}^{(\times)}=1$, $e_{ij}^{(+)}e_{ij}^{(\times)}=0$.

From the Einstein equation, the equation of motion of GWs at second order is given by
\begin{equation}
	h_{ij}''+2\mathcal Hh_{ij}'-\nabla^2h_{ij}=-4\hat{\mathcal T}_{ij}{}^{lm}\mathcal S_{lm},\label{EOM_GW1}
\end{equation}
where $'$ indicates the derivative with respect to the conformal time $\eta$, $\mathcal H=a'/a$ is the conformal Hubble parameter, $\mathcal S_{lm}$ is  the source term given below and
$\hat{\mathcal T}_{ij}{}^{lm}$ is the projection operator onto the transverse and traceless tensor which is expressed in the  Fourier space as
\begin{equation}
	{\mathcal T}_{ij}{}^{lm}(\bvec k)=e_{ij}^{(+)}(\bvec k)e^{(+)lm}(\bvec k)+e_{ij}^{(\times)}(\bvec k)e^{(\times)lm}(\bvec k).
\end{equation}
The source term is expanded in a Fourier series as 
\begin{equation}
	\mathcal S_{lm}(\eta,\bvec x)=\int\frac{\mathrm d^3k}{(2\pi)^3}\tilde{\mathcal S}_{lm}(\eta,\bvec k)e^{i\bvec k\cdot \bvec x}.
\end{equation}
The equation of motion Eq.~\eqref{EOM_GW1} is rewritten in terms of each polarization mode as
\begin{equation}
	{h_{\bvec k}^s}''(\eta)+2\mathcal H{h_{\bvec k}^s}'(\eta)+k^2h_{\bvec k}^s(\eta)=\label{EOM_GW2}\mathcal S^s(\eta,\bvec k),
\end{equation}
where $s=(+),\,(\times)$ and $\mathcal S^s(\eta,\bvec k)\equiv-4e^{s,lm}(\bvec k)\mathcal S_{lm}(\eta,\bvec k)$.
The solution of Eq.~\eqref{EOM_GW2} in the radiation-dominated universe is given by
\begin{equation}
	h_{\bvec k}^s(\eta)=\frac{1}{a(\eta)}\int^\eta\mathrm d\bar\eta g_{\bvec k}(\eta;\bar\eta)a(\bar\eta)\mathcal S^s(\bar\eta,\bvec k),\label{h_solution}
\end{equation}
where
\begin{equation}
	g_{\bvec k}(\eta;\bar\eta)=\frac{\sin\left[k(\eta-\bar\eta)\right]}{k}\theta(\eta-\bar\eta)\label{Green}
\end{equation}
is the Green function and $\theta(\eta)$ is the step function.

\subsection{Source Term}

The source term $\mathcal S_{ij}$ in Eq.~\eqref{EOM_GW1} coming from the second order of the scalar metric perturbation is given by~\cite{Acquaviva:2002ud}\footnote{
There is also the source term coming from the kinetic term of the curvaton, but the contribution of it is subdominant compared with that of the scalar metric perturbation.~\cite{Kawasaki:2013xsa}
}
\begin{equation}
	\mathcal S_{ij}=4\Psi\partial_i\partial_j\Psi+2\partial_i\Psi\partial_j\Psi-\frac{4}{3(1+w)}\partial_i\left(\frac{\Psi'}{\mathcal H}+\Psi\right)\partial_j\left(\frac{\Psi'}{\mathcal H}+\Psi\right).
\end{equation}
Hereafter, we take $w=1/3$. Then, the right-hand side of Eq.~\eqref{EOM_GW2} is calculated as
\begin{eqnarray}
	\mathcal S^s(\eta,\bvec k)=4\int\frac{\mathrm d^3q}{(2\pi)^3}e^{s,ij}(\bvec k)q_iq_j\left[3\tilde\Psi(\bvec q)\tilde\Psi(\bvec k-\bvec q)
	+\frac{2}{\mathcal H}\tilde\Psi'(\bvec q)\tilde\Psi(\bvec k-\bvec q)+\frac{1}{\mathcal H^2}\tilde\Psi'(\bvec q)\tilde\Psi'(\bvec k-\bvec q)\right].\label{source}
\end{eqnarray}
Here, the Fourier mode of the scalar metric perturbation is defined as
\begin{equation}
	\tilde\Psi(\eta,\bvec k)=\int\mathrm d^3x\Psi(\eta,\bvec x)e^{-i\bvec k\cdot\bvec x}
\end{equation}
and the $\eta$ dependence is made implicit in Eq.~\eqref{source}. In deriving Eq.~\eqref{source}, it is important to note that
\begin{equation}
	\int \mathrm d^3qe^{s,ij}(\bvec k)q_iq_j\tilde\Psi'(\bvec k-\bvec q)\tilde\Psi(\bvec q)=
	\int \mathrm d^3\tilde qe^{s,ij}(\bvec k)(k_i-\tilde q_i)(k_j-\tilde q_j)\tilde\Psi'(\tilde{\bvec q})\tilde\Psi(\bvec k-\tilde{\bvec q})=
	\int \mathrm d^3\tilde qe^{s,ij}(\bvec k)\tilde q_i\tilde q_j\tilde\Psi'(\tilde{\bvec q})\tilde\Psi(\bvec k-\tilde{\bvec q})
\end{equation}
since $e^{s,ij}(\bvec k)$ is transverse.

Now, let us discuss the source term emerging in the axion-like curvaton model. It is useful to represent the time dependence of $\tilde\Psi(\eta,\bvec k)$ in terms of a transfer function as\footnote{
See Footnote~\ref{magnitude_k}.
}
\begin{equation}
	\tilde\Psi(\eta,\bvec k)=T(\eta,k)S(\bvec k),\label{trans}
\end{equation}
where $S$ is the isocurvature perturbation induced by the curvaton and is defined by Eq.~\eqref{isocurvature}.
From Eqs.~\eqref{source} and \eqref{trans}, the source term is rewritten as
\begin{equation}
	\mathcal S^s(\eta,\bvec k)=\int\frac{\mathrm d^3q}{(2\pi)^3}e^{s}(\bvec k,\bvec q)f(q,|\bvec k-\bvec q|,\eta)S(\bvec q)S(\bvec k-\bvec q),\label{source2}
\end{equation}
where
\begin{equation}
	e^{s}(\bvec k,\bvec q)\equiv e^{s,ij}(\bvec k)q_iq_j=
		\begin{cases}
			\frac{1}{\sqrt 2}q^2\sin^2\theta\cos 2\phi\quad\mathrm{for}\quad s=(+)\\
			\frac{1}{\sqrt 2}q^2\sin^2\theta\sin 2\phi\quad\mathrm{for}\quad s=(\times)\\
		\end{cases}\label{spherical}
\end{equation}
and
\begin{equation}
	f(k_1,k_2,\eta)\equiv 4\left[3T(k_1)T(k_2)
	+\frac{2}{\mathcal H}T'(k_1)T(k_2)+\frac{1}{\mathcal H^2}T'(k_1)T'(k_2)\right].\label{f}
\end{equation}
In Eq.\eqref{spherical}, we take $\bvec e(\bvec k)$, $\bar{\bvec e}(\bvec k)$ and $\bvec k$ as $x$, $y$ and $z$ direction respectively, and use a spherical coordinate $(q,\theta,\phi)$. In Eq. \eqref{f}, the $\eta$ dependence of $T(\eta,k)$ is made implicit. We mention that there are some symmetries such as $e^{s}(\bvec k,\bvec q)=e^{s}(-\bvec k,\bvec q)=e^{s}(\bvec k,-\bvec q)$ and that we can take $f(k_1,k_2,\eta)=f(k_2,k_1,\eta)$ in the integral. In Eq. \eqref{source2}, $f(k_1,k_2,\eta)$ can be replaced by 
\begin{equation}
	\bar f(k_1,k_2,\eta)\equiv 4\left[2T(k_1)T(k_2)
	+\left(T(k_1)+\frac{T'(k_1)}{\mathcal H}\right)\left(T(k_2)+\frac{T'(k_2)}{\mathcal H}\right)\right].
\end{equation}
This function is symmetric under $k_1\leftrightarrow k_2$ explicitly. In terms of $f(k_1,k_2,\eta)$, the same symmetry goes on only in the integral.

\subsection{Power Spectrum}

From Eqs.\eqref{h_solution} and \eqref{source2}, the two-point function of each polarization mode of GWs is written as
\begin{eqnarray}
	\left<h_{\bvec k}^r(\eta)h_{\bvec p}^s(\eta)\right>&=&
	\frac{1}{a^2(\eta)}\int^\eta\mathrm d\bar\eta_1a(\bar\eta_1)g_{\bvec k}(\eta;\bar\eta_1)\int^\eta\mathrm d\bar\eta_2a(\bar\eta_2)g_{\bvec p}(\eta;\bar\eta_2)\left<\mathcal S^r(\bvec k,\bar\eta_1)\mathcal S^s(\bvec p,\bar\eta_2)\right>\\
	&=&\frac{1}{a^2(\eta)}\int\frac{\mathrm d^3\tilde k}{(2\pi)^3}\int\frac{\mathrm d^3\tilde p}{(2\pi)^3}e^{r}(\bvec k,\bvec{\tilde k})e^{s}(\bvec p,\bvec{\tilde p})\left<S(\bvec{\tilde k})S(\bvec k-\tilde{\bvec k})S(\bvec{\tilde p})S(\bvec p-\tilde{\bvec p})\right>\nonumber\\
	&&\times\int^\eta\mathrm d\bar\eta_1a(\bar\eta_1)g_{\bvec k}(\eta;\bar\eta_1)f(\tilde k,|\bvec k-\bvec{\tilde k}|,\bar\eta_1)
	\int^\eta\mathrm d\bar\eta_2a(\bar\eta_2)g_{\bvec p}(\eta;\bar\eta_2)f(\tilde p,|\bvec p-\bvec{\tilde p}|,\bar\eta_2)\label{2point_h}.
\end{eqnarray}
The four-point function of the isocurvature perturbations $\left<SSSS\right>$ is evaluated as the sum of three configurations of contraction. Neglecting the zero momentum mode, we obtain
\begin{equation}
	\left<S(\bvec{\tilde k})S(\bvec k-\tilde{\bvec k})S(\bvec{\tilde p})S(\bvec p-\tilde{\bvec p})\right>=(2\pi)^6\delta(\bvec k+\bvec p)\left[\delta(\bvec{\tilde k}+\bvec{\tilde p})+\delta(\bvec k-\bvec{\tilde k}+\bvec{\tilde p})\right]\frac{2\pi^2}{\tilde k^3}\frac{2\pi^2}{|\bvec k-\tilde{\bvec k}|^3}\mathcal P_S(\tilde k)\mathcal P_S(|\bvec k-\tilde{\bvec k}|), \label{4point_S}
\end{equation}
where the power spectrum of the isocurvature perturbations is defined by
\begin{equation}
	\left<S(\bvec k)S(\bvec p)\right>\equiv(2\pi)^3\delta(\bvec k+\bvec p)\frac{2\pi^2}{k^3}\mathcal P_S(k).
\end{equation}
The contributions of the two terms in the brackets in Eq.~\eqref{4point_S} to Eq.~\eqref{2point_h} are the same. It is realized by erasing $\bvec p$, $\bvec{\tilde p}$ using the delta function and taking into account the symmetries mentioned at the end of the previous subsection. Then, Eq.~\eqref{2point_h} becomes
\begin{eqnarray}
	\left<h_{\bvec k}^r(\eta)h_{\bvec p}^s(\eta)\right>=2(2\pi)^3\delta(\bvec k+\bvec p)\frac{1}{a^2(\eta)}&&\int\frac{\mathrm d^3\tilde k}{(2\pi)^3}e^{r}(\bvec k,\bvec{\tilde k})e^{s}(\bvec k,\bvec{\tilde k})\frac{2\pi^2}{\tilde k^3}\frac{2\pi^2}{|\bvec k-\tilde{\bvec k}|^3}\mathcal P_S(\tilde k)\mathcal P_S(|\bvec k-\tilde{\bvec k}|)\nonumber\\
	&&\quad\times\left[\int^\eta\mathrm d\bar\eta a(\bar\eta)g_{\bvec k}(\eta;\bar\eta)f(\tilde k,|\bvec k-\bvec{\tilde k}|,\bar\eta)\right]^2.
\end{eqnarray}
After implementing $\phi$ integral using Eq.~\eqref{spherical}, it is realized that
\begin{eqnarray}
	\left<h_{\bvec k}^{(+)}(\eta)h_{\bvec p}^{(\times)}(\eta)\right>&=&0 \\
	\left<h_{\bvec k}^{(+)}(\eta)h_{\bvec p}^{(+)}(\eta)\right>&=&\left<h_{\bvec k}^{(\times)}(\eta)h_{\bvec p}^{(\times)}(\eta)\right>\nonumber\\
	&=&(2\pi)^3\delta(\bvec k+\bvec p)\frac{\pi^2}{2}\int_0^\infty\mathrm d\tilde k\int_{-1}^1\mathrm d\mu(1-\mu^2)^2\frac{\tilde k^3}{|\bvec k-\tilde{\bvec k}|^3}\mathcal P_S(\tilde k)\mathcal P_S(|\bvec k-\tilde{\bvec k}|)\nonumber\\
	&&\qquad\times\left[\frac{1}{a(\eta)}\int^\eta\mathrm d\bar\eta a(\bar\eta)g_{\bvec k}(\eta;\bar\eta)f(\tilde k,|\bvec k-\bvec{\tilde k}|,\bar\eta)\right]^2,\label{2point_h_2}
\end{eqnarray}
where we defined $\mu\equiv\cos \theta$. The power spectrum of the GWs is defined by
\begin{equation}
	\left<h_{\bvec k}^r(\eta)h_{\bvec p}^s(\eta)\right>\equiv(2\pi)^3\delta(\bvec k+\bvec p)\delta^{rs}\frac{2\pi^2}{k^3}\mathcal P_h(\eta,k).\label{P_h_def}
\end{equation}
From Eqs.~\eqref{2point_h_2} and \eqref{P_h_def}, the power spectrum is obtained as
\begin{eqnarray}
	\mathcal P_h(\eta,k)&=&\frac{1}{4}\int_0^\infty dy\int_{|1-y|}^{1+y}dx\frac{y^2}{x^2}\left(1-\frac{(1+y^2-x^2)^2}{4y^2}\right)^2\mathcal P_S(kx)\mathcal P_S(ky)
	\left[\frac{k^2}{a(\eta)}\int^\eta d\tilde\eta a(\bar\eta)g_{\bvec k}(\eta;\bar\eta)f(ky,kx,\bar\eta)\right]^2
	,\label{P_h}
\end{eqnarray}
where we defined $x\equiv|\bvec k-\tilde{\bvec k}|/k$ and $y\equiv\tilde k/k$. 
Note that $x^2=1+y^2-2y\mu$. Now, the power spectrum can be calculated numerically by using Eq.~\eqref{P_h} because $g_{\bvec k}(\eta;\bar\eta)$ and $f(k_1,k_2,\eta)$ are given by Eqs.~\eqref{Green} and \eqref{f}. Moreover, the energy density of GWs can be calculated with the power spectrum as shown in the next subsection.

\subsection{Energy Density}

The energy density of the GWs is given by~\cite{Maggiore:1999vm}
\begin{equation}
	\rho_\mathrm{GW}=\frac{M_P^2}{16a^2}\left<\frac{1}{2}\overline{\left(h_{ij}'\right)^2}+\frac{1}{2}\overline{\left(\nabla h_{ij}\right)^2}\right>\simeq\frac{M_P^2}{16a^2}\left<\overline{\left(\nabla h_{ij}\right)^2}\right>,
\end{equation}
where the overlines stand for the time average. Furthermore, the energy density is rewritten in terms of the power spectrum as
\begin{eqnarray}
	\rho_\mathrm{GW}(\eta)&=&\int\mathrm d\ln k\,\rho_\mathrm{GW}(\eta,k)\\
	\rho_\mathrm{GW}(\eta,k)&\equiv&\frac{M_P^2}{8}\left(\frac{k}{a}\right)^2\overline{\mathcal P_h(\eta,k)},
\end{eqnarray}
which leads to the density parameter of the GWs within the logarithmic interval of the wave number,
\begin{equation}
	\Omega_{\mathrm{GW}}(\eta,k)\equiv\frac{\rho_\mathrm{GW}(\eta,k)}{\rho_\mathrm{cr}(\eta)}=\frac{1}{24}\left(\frac{k}{aH}\right)^2\overline{\mathcal P_h(\eta,k)}.
\end{equation}
The formula for the tensor power spectrum given in Eq.~\eqref{P_h} is valid only in the radiation-dominated universe.
After the wavelength reenters the horizon, the amplitude of $h$ is inversely proportion to the scale factor $a(\eta)$ and the energy density of the GWs dilutes in the same way as the radiation.
Then, the current density parameter of the GWs is obtained as
\begin{eqnarray}
	 \Omega_{\mathrm{GW}}(\eta_0,k)
	 &=&\prn{\frac{ a_\star^2H_\star    }{ a_0^2H_0  }}^2\Omega_{\mathrm{GW}}(\eta_\star,k)
	 =0.83\prn{ \frac{g_{*,\star}}{10.75}    }^{-1/3}\Omega_{r,0}\Omega_{\mathrm{GW}}(\eta_\star,k)\nonumber\\
	&=&0.83\prn{ \frac{g_{*,\star}}{10.75}    }^{-1/3}\frac{\Omega_{r,0}k^2}{24\mathcal H(\eta_\star)^2}\overline{\mathcal P_h(\eta_\star,k)},
	 \label{Omega_GW}
\end{eqnarray}
where ``$\star$" represents the value soon after the GWs begin to behave as radiation
and $\Omega_{r,0}\simeq9.1\times10^{-5}$ is the current density parameter of the radiation assuming that neutrinos are relativistic.

\section{Power Spectrum with Non-Gaussianity}

In this appendix, we review the analysis of quadratic non-Gaussianity. Especially, we see the case where the power spectrum of the Gaussian part is scale-invariant as a simple example. In Sec.~\ref{subsec:NG_suppression}, the formalism is applied to the axion-like curvaton model.

A curvature perturbation with quadratic non-Gaussianity is expressed using $f_\mathrm{NL}$ as
\begin{equation}
	\zeta(\bvec x)=\zeta_g(\bvec x)+\frac{3}{5}f_\mathrm{NL}\left(\zeta_g^2(\bvec x)-\left<\zeta_g^2(\bvec x)\right>\right),\label{quadratic_NG}
\end{equation}
where $\zeta_g(\bvec x)$ follows a Gaussian distribution. The factor $\frac{3}{5}$ comes from the fact that the relation of curvature perturbation, $\zeta$, and the curvature perturbation in the conformal Newtonian gauge, $\Psi$, is given by $\Psi=-\frac{3}{5}\zeta$ on super-horizon scales in the matter-dominated universe. 
We use the following convention for Fourier transformation:
\begin{equation}
	\begin{cases}
		f(\bvec x)=\int\frac{\mathrm d^3k}{(2\pi)^3}\tilde f(\bvec k)e^{i\bvec k\cdot \bvec x}\\
		\tilde f(\bvec k)=\int \mathrm d^3x f(\bvec x)e^{-i\bvec k\cdot \bvec x}
	\end{cases}.
\end{equation}
If we define the nonlinear part as
\begin{equation}
	\zeta_\mathrm{NL}(\bvec x)\equiv\zeta_g^2(\bvec x)-\left<\zeta_g^2(\bvec x)\right>,
\end{equation}
the Fourier mode of Eq.~\eqref{quadratic_NG} is written as
\begin{equation}
	\tilde\zeta(\bvec k)=\tilde\zeta_g(\bvec k)+\frac{3}{5}f_\mathrm{NL}\tilde\zeta_\mathrm{NL}(\bvec k).
\end{equation}
$\tilde\zeta_\mathrm{NL}(\bvec k)$ is calculated straightforwardly as
\begin{eqnarray}
	\tilde\zeta_\mathrm{NL}(\bvec k)&=&\int\mathrm d^3x\left(\zeta_g^2(\bvec x)-\left<\zeta_g^2(\bvec x)\right>\right)e^{-i\bvec k\cdot\bvec x}\\
	&=&\int\frac{\mathrm d^3p}{(2\pi)^3}\tilde\zeta_g(\bvec p)\tilde\zeta_g(\bvec k-\bvec p)-(2\pi)^3\delta(\bvec k)\left<\zeta_g^2(\bvec x)\right>.\label{Spergel}
\end{eqnarray}
We mention that the first term has other expressions: $\int\frac{\mathrm d^3p}{(2\pi)^3}\tilde\zeta_g(\bvec p)\tilde\zeta_g(\bvec k-\bvec p)=\int\frac{\mathrm d^3p}{(2\pi)^3}\tilde\zeta_g(\bvec p+\bvec k)\tilde\zeta_g(-\bvec p)=\int\frac{\mathrm d^3p}{(2\pi)^3}\tilde\zeta_g(\bvec p+\bvec k)\tilde\zeta_g^*(\bvec p)$.\\
\indent The correlation function of $\zeta$ in the real space is
\begin{equation}
	\left<\zeta^2(\bvec x)\right>=\left<\zeta_g^2(\bvec x)\right>+\left(\frac{3}{5}f_\mathrm{NL}\right)^2\left<\zeta_\mathrm{NL}^2(\bvec x)\right>\label{zeta_x1}
\end{equation}
where
\begin{equation}
	\left<\zeta_\mathrm{NL}^2(\bvec x)\right>=\left<\zeta_g^4(\bvec x)\right>-\left<\zeta_g^2(\bvec x)\right>^2=2\left<\zeta_g^2(\bvec x)\right>^2.\label{zeta_x2}
\end{equation}
We have used the formula for the four-point function of the Gaussian variable: $\left<\zeta_g^4(\bvec x)\right>=3\left<\zeta_g^2(\bvec x)\right>^2$ (the factor 3 can be interpreted as the number of patterns of contraction). Combining Eqs.~\eqref{zeta_x1} and \eqref{zeta_x2}, we obtain
\begin{equation}
	\left<\zeta^2(\bvec x)\right>=\left<\zeta_g^2(\bvec x)\right>+2\left(\frac{3}{5}f_\mathrm{NL}\right)^2\left<\zeta_g^2(\bvec x)\right>^2\label{zeta_x3}.
\end{equation}
Now, we will obtain a corresponding expression in the momentum space and check that it reproduces Eq.~\eqref{zeta_x3}.\\
\indent The power spectrum of some quantity $f$ is defined by
\begin{eqnarray}
	\left<\tilde f(\bvec k)\tilde f(\bvec p)\right>&\equiv&(2\pi)^3\delta(\bvec k+\bvec p)P_f(k)\\
	\mathcal P_f(k)&\equiv&\frac{k^3}{2\pi^2}P_f(k).
\end{eqnarray}
Then, it can be shown that $\left<f(\bvec x)f(\bvec y)\right>=\int\frac{\mathrm d^3k}{(2\pi)^3}P_f(k)e^{i\bvec k\cdot(\bvec x-\bvec y)}=\int\frac{\mathrm dk}{k}\mathcal P_f(k)\frac{\sin\left(k|\bvec x-\bvec y|\right)}{k|\bvec x-\bvec y|}$ and $\left<\tilde\zeta_\mathrm{NL}(\bvec k)\right>=0$ because of Eq.~\eqref{Spergel}.
Taking into account that the correlation function of an odd number of $\tilde\zeta_g$ is zero, we derive
\begin{equation}
	\left<\tilde \zeta(\bvec k)\tilde \zeta(\bvec p)\right>=\left<\tilde \zeta_g(\bvec k)\tilde \zeta_g(\bvec p)\right>+\left(\frac{3}{5}f_\mathrm{NL}\right)^2\left<\tilde \zeta_\mathrm{NL}(\bvec k)\tilde \zeta_\mathrm{NL}(\bvec p)\right>.\label{2PointFunc}
\end{equation}
Thanks to Eq.~\eqref{Spergel}, the last term is evaluated as
\begin{eqnarray} 
	\left<\tilde \zeta_\mathrm{NL}(\bvec k)\tilde \zeta_\mathrm{NL}(\bvec p)\right>=&&\int\frac{\mathrm d^3q_1}{(2\pi)^3}\int\frac{\mathrm d^3q_2}{(2\pi)^3}\left<\tilde\zeta_g(\bvec q_1)\tilde\zeta_g(\bvec k-\bvec q_1)\tilde\zeta_g(\bvec q_2)\tilde\zeta_g(\bvec p-\bvec q_2)\right>\nonumber\\
	&&-(2\pi)^3\delta(\bvec k)\left<\zeta_g^2(\bvec x)\right>\int\frac{\mathrm d^3q}{(2\pi)^3}\left<\tilde\zeta_g(\bvec q)\tilde\zeta_g(\bvec p-\bvec q)\right>+(\bvec k\leftrightarrow\bvec p)\nonumber\\
	&&+(2\pi)^6\delta(\bvec k)\delta(\bvec p)\left<\zeta_g^2(\bvec x)\right>^2\label{NLNL}.
\end{eqnarray}
The four-point function is evaluated by summing over three patterns of contraction as
\begin{equation}\begin{split}
	\int\frac{\mathrm d^3q_1}{(2\pi)^3}&\int\frac{\mathrm d^3q_2}{(2\pi)^3}\left<\tilde\zeta_g(\bvec q_1)\tilde\zeta_g(\bvec k-\bvec q_1)\tilde\zeta_g(\bvec q_2)\tilde\zeta_g(\bvec p-\bvec q_2)\right>\\
	&=(2\pi)^6\delta(\bvec k)\delta(\bvec p)\left<\zeta_g^2(\bvec x)\right>^2+2(2\pi)^3\delta(\bvec k+\bvec p)\int\frac{\mathrm d^3q}{(2\pi)^3}P_{\zeta_g}(\bvec q)P_{\zeta_g}(|\bvec k-\bvec q|).
\end{split}\end{equation}
The terms containing $\left<\zeta_g^2(\bvec x)\right>^2$ in Eq.~\eqref{NLNL} are cancelled out beautifully. 
Therefore, we derive
\begin{equation}
	\left<\tilde \zeta_\mathrm{NL}(\bvec k)\tilde \zeta_\mathrm{NL}(\bvec p)\right>=2(2\pi)^3\delta(\bvec k+\bvec p)\int\frac{\mathrm d^3q}{(2\pi)^3}P_{\zeta_g}(\bvec q)P_{\zeta_g}(|\bvec k-\bvec q|),
\end{equation}
which is rewritten as
\begin{eqnarray}
	P_{\zeta_\mathrm{NL}}(k)&=&2\int\frac{\mathrm d^3p}{(2\pi)^3}P_{\zeta_g}(\bvec p)P_{\zeta_g}(|\bvec k-\bvec p|)\\
	\mathcal P_{\zeta_\mathrm{NL}}(k)&=&\frac{k^3}{2\pi}\int\mathrm d^3p\frac{1}{p^3}\frac{1}{\left|\bvec k-\bvec p\right|^3}\mathcal P_{\zeta_g}(p)\mathcal P_{\zeta_g}(|\bvec k-\bvec p|).\label{PNL_clean}
\end{eqnarray}

Hereafter, we assume that the power spectrum is scale-invariant. Moreover, when the IR cutoff of the integral is $L^{-1}\simeq H$, we assume that $kL \gg 1$. 
The IR cutoff means that the integrand is zero inside the sphere of radius $L^{-1}$ around each pole. The integral in Eq.~\eqref{PNL_clean} is evaluated as
\begin{eqnarray}
	\frac{k^3}{2\pi}\int\mathrm d^3p\frac{1}{p^3}\frac{1}{\left|\bvec k-\bvec p\right|^3}&=&k^3\int\mathrm dp\int_{-1}^1\mathrm d(\cos\theta)\frac{1}{p^3}\frac{1}{(p^2+k^2-2pk\cos\theta)^{3/2}}\nonumber\\
	&=&k^2\int\mathrm dp\frac{1}{p^2}\left(\frac{1}{|p-k|}-\frac{1}{p+k}\right)\nonumber\\
	&=&k^2\left[\left(\int_{k+L^{-1}}^\infty-\int_{L^{-1}}^{k-L^{-1}}\right)\frac{\mathrm dp}{p^2(p-k)}-\int_{L^{-1}}^\infty\frac{\mathrm dp}{p^2(p+k)}\right]\nonumber\\
	&=&-2+2\ln (kL+1)+2\ln(kL-1)\nonumber\\
	&\simeq&4\ln(kL).
\end{eqnarray}
Therefore, under the assumptions mentioned above, Eq.~\eqref{PNL_clean} becomes
\begin{equation}
	\mathcal P_{\zeta_\mathrm{NL}}(k)=4\mathcal P_{\zeta_g}^2\ln(kL),\label{factor4}
\end{equation}
and so Eq.~\eqref{2PointFunc} leads to
\begin{equation}
	\mathcal P_\zeta(k)=\mathcal P_{\zeta_g}+4\left(\frac{3}{5}f_\mathrm{NL}\right)^2 \mathcal P_{\zeta_g}^2\ln(kL).\label{zeta_k}
\end{equation}
\indent One can check that Eq.~\eqref{factor4} reproduces the correct expression in the real space, Eq.~\eqref{zeta_x2}. If the UV cutoff of the momentum integral is denoted as $k_\mathrm{max}$, $\left<\zeta_g^2(\bvec x)\right>$ is expressed as
\begin{equation}
	\left<\zeta_g^2(\bvec x)\right>=\int_{L^{-1}}^{k_\mathrm{max}}\frac{\mathrm dk}{k}\mathcal P_{\zeta_g}=\mathcal P_{\zeta_g}\ln(k_\mathrm{max}L).\label{P_zeta_g_log}
\end{equation}
Similarly, $\left<\zeta_\mathrm{NL}^2(\bvec x)\right>$ is calculated as
\begin{equation}
	\left<\zeta_\mathrm{NL}^2(\bvec x)\right>=\int_{L^{-1}}^{k_\mathrm{max}}\frac{\mathrm dk}{k}\mathcal P_{\zeta_\mathrm{NL}}(k)=4\mathcal P_{\zeta_g}^2\int_{L^{-1}}^{k_\mathrm{max}}\frac{\mathrm dk}{k}\ln(kL)=2\mathcal P_{\zeta_g}^2\left(\ln(k_\mathrm{max}L)\right)^2=2\left<\zeta_g^2(\bvec x)\right>^2.
\end{equation}
This is exactly Eq.~\eqref{zeta_x2}.

\small
\bibliographystyle{apsrev4-1}


%


\end{document}